\documentstyle[12pt,amssymb,aps,epsfig,prb]{revtex}
\tightenlines
\begin{document}
\begin{titlepage}
\title{Electric field effect on positronium formation in liquids.}
\author{Sergey V.~Stepanov and Vsevolod M.~Byakov \\
        Institute of Theoretical and Experimental Physics,
        Moscow 117218, Russia }
\maketitle

\vspace{5mm}
\begin{abstract}
The comparison of different models (the Ore, spur and blob models) of positronium (Ps) formation is presented.  Because in
molecular media Ps is formed in the terminal positron blob and not in an ordinary spur, the application of the blob model seems
to be the most adequate. We extend this model for consideration of the Ps formation in the presence of external electric field
($<100$ kV/cm).  In the simplified limiting case, this approach provides a formula similar to the Onsager one for the geminate
recombination probability. The influence of ion-electron recombination and other intrablob processes on Ps formation is taken
into account. The role of quasifree positronium in Ps formation process is discussed.
\end{abstract}

\vspace{5mm}
PACS number (s): 78.70.Bj, 82.55.+e, 36.10.Dr

\vspace{5mm}
The author to whom all correspondence should be sent:\\
{\bf Dr. Sergey V. Stepanov} \\
Laboratory of Neutron Physics \\
Institute of Theoretical and Experimental Physics \\
Bolshaya Cheremushkinskaya st., 25, \\
Moscow, 117218 \\
RUSSIA

\medskip
\noindent
e-mail: Sergey.Stepanov@itep.ru  \qquad \qquad fax: 7-095-125-7124

\end{titlepage}

\section{Introduction}

Interpretation of the data obtained by means of different methods of the positron annihilation
spectroscopy (PAS) is largely based on theoretical understanding of primary physico-chemical
reactions and processes in the terminal part of the positron track, and on comprehension of the
nature of the states from which the positrons may annihilate. \cite{Mog95,Bya85}

Because of extreme rapidity of the primary radiation chemical processes, fundamental knowledge about them is
obtained mainly from theoretical analysis and interpretation of relevant experimental data and first of all
from pulse radiolysis, permitting to enter the area of pico- and (probably, in the near future) femtosecond
time scale.\cite{Tag00} PAS and especially the experiments in external electric fields allow one to obtain
important supplementary information about the intratrack processes, namely ion-electron recombination, e$^+$
and e$^-$ thermalization and localization, positronium (Ps) formation, influence of various chemical
additives and to make conclusions about the spatial structure of the e$^+$ track. \cite{Ito88} These advances
throughout the last two decades became possible due to essential instrumental achievements (application of
low-energy positron beams, better time resolution, accumulation of higher statistics) and theoretical
development (calculations of cross sections and rate constants of a variety of fundamental processes with
participation of positrons, investigations of e$^+$-e$^-$-correlations, slowing down, track effects). This
progress significantly increased the reliability of PAS.

First experiments on Ps formation (including those in external electric field) were carried out
in low-density simple atomic gases, where Ps is formed predominantly according to the Ore
mechanism. \cite{Teu56,Mar56} It implies that the "hot" positron, e$^{+*}$, pulls out an electron
from a molecule M, forming Ps atom and leaving behind the radical-cation M$^{+\cdot}$:

$$
\rm e^{+*}+M \to Ps + M^{+\cdot}.
\eqno(1)
$$
This process is the most effective, when the energy $W$ of the positron lays within the interval named the
Ore gap:

$$
I_G-Ry/2 < W < W_{ex}. \eqno(2)
$$
Here $I_G$ is the first ionization potential of the molecule, $W_{ex}$ is its electronic excitation threshold
and $Ry/2=6.8$ eV is the Ps binding energy in vacuum. It is believed, that the positron with the energy lower
than $I_G-Ry/2$ cannot pick up an electron from a molecule. When $W>W_{ex}$ electronic excitations and
ionizations compete with Ps formation and the last one becomes less effective.

Theoretical consideration of the electric field  effect on Ps formation in gases was based on a solution of the Boltzmann
kinetic equation. \cite{Teu56} This approach successfully explained the rather sharp growth of the Ps yield vs the field in
rare gases, molecular hydrogen (H$_2$, D$_2$) and N$_2$ at moderate densities. \cite{Mar56} Electric field hinders energy loss
by the positrons and at the same time "heats" them up to the energies above the lower boundary of the Ore gap. The simple
estimation of the field $\bf E$ needed to keep e$^+$ within the Ore gap can be obtained if we equate a gain of the e$^+$
kinetic energy between two subsequent collisions

$$
\left\langle
    \frac{m}{2}\left( {\textbf{\textit{v}}} +
\frac{e{\bf E} l_{tr}}{mv} \right)^2 - \frac{mv^2}{2} \right\rangle_v
                                =
        \frac{(e E l_{tr})^2}{2mv^2}
\eqno(3)
$$
and the average energy loss, $\frac{2m}{M}\frac{mv^2}{2}$ (this expression corresponds to the energy loss in
elastic collisions). Here $m$ is the e$^+$ mass, $M$ is the mass of a gas molecule and $l_{tr}$ is the
positron transport mean free path (Appendix A). Averaging $\langle \dots \rangle_v$ is carried out over
orientations of the velocity {\textbf {\textit {v}}} of the positron. From such a balance condition one
obtains that the average e$^+$ kinetic energy $\langle W \rangle_v$ is approximately equal to
$(M/m)^{1/2}eEl_{tr}$. At typical gas densities and external fields 10-500 V/cm it falls in the eV region.
Thus, rather moderate electric fields are able to accelerate positrons up to the energies needed for
realization of reaction (1). It is worth mentioning that in liquid helium the field dependence of the Ps
formation probability, $P_{\rm Ps}$, is qualitatively the same as in the gas phase. \cite{Man66} It means
that the Ore process contributes to the Ps formation in liquid helium. Respective growth of the $P_{\rm Ps}$
occurs in fields, approximately 500 times larger than in the gas phase, which is in accordance with the above
estimation. This increase of the threshold field is determined by the decrease of $l_{tr}$, which is
inversely proportional to the density of helium.

More detailed theoretical considerations of Ps formation problem in light noble gases require a knowledge of energy
dependencies of momentum transfer, $\sigma_m$, and annihilation, $\sigma_a$, cross sections of e$^+$ vs its kinetic energy (see
Ref.\cite{Shi87} and references therein). In terms of energy dependent cross sections the theory succeeded to explain the most
striking experimental observation in gases:  so-called "shoulder" in the e$^+$ lifetime annihilation spectra. \cite{Cha85} It
appears due to existence of the Ramsauer minimum in $\sigma_m(W)$ which usually takes place at $W\sim 1$ eV.

In the 1990's, the Ps formation in noble gases was studied over a wide range of densities in the presence of static electric
field. \cite{Cha92,Pep95} One of the motivations for these studies was a significant contradiction between available
experimental values of $P_{\rm Ps}$ in heavier noble gases (Kr, Xe) and corresponding predictions of the standard Ore model.
Another unanswered question was why the application of a strong electric field does not increase $P_{\rm Ps}$ up to unity?
Probably the origin of the effects (as well as a surprising decrease of $P_{\rm Ps}$ at higher fields in Xe) is related to the
formation of localized states (due to the self-trapping) by both e$^+$ and Ps prior to annihilation. \cite{Iak82}

In molecular gases (H$_2$, N$_2$, CO$_2$, CH$_4$) with low-lying vibrational or rotational levels, Ps yield reveals complex
behavior with changes in the density, temperature and electric field. \cite{Cha85} It cannot be convincingly treated in terms
of e$^+$-M and Ps-M cross sections. However experimental data can be fitted by means of combination of the Ore process and
recombination mechanism. The last one suggests Ps formation in condensed media through combination of the thermalized positron
with one of track electrons produced by e$^+$ ionization of the medium in the terminal part of the e$^+$
track.\cite{Mog74,Bya74,Bya76} Later on, the Onsager-like formulation of the recombination mechanism (the spur model
\cite{Mog74}) became very widespread in spite of ignorance of the presence of positive molecular ions and other intrablob
electrons, different mobilities of e$^+$ and e$^-$ (Sec. II). Anyhow, combination of the Ore model and the spur model
satisfactorily reproduces experimentally observed variations of $P_{\rm Ps}$ vs density and electric field.
\cite{Cur85,Jac86,Pep93}

Since the mid-1970's the Ps formation at low fields ($<200$ kV/cm) was also investigated in liquid and solid paraffins
\cite{Bra75,Wan01} and some pure liquids. \cite{Ani75,Lin85,Wan98} Below several tens of kV/cm the Ps yield noticeably drops,
then the falling decelerates and a tendency to reach a plateau is observed. Application of much stronger fields ($\sim 1$
MV/cm) may push the positron in the Ore gap and keep it there until the Ps will be formed, or e$^+$ annihilate with one of the
molecular electrons. Thereby the strong electric field stimulates reaction (1). It is very probable that this phenomenon was
experimentally observed in solid (polar and nonpolar) polymers \cite{Bis83}, though the other factors (formation of bulk and
surface charges, field induced electron and positron trapping/detrapping on structural defects) were mentioned, which may also
increase the Ps yield in the fields about 1 MV/cm. \cite{Mog83}

A theoretical interpretation of these experiments in condensed phase was attempted by Brandt et al.
\cite{Bra68,Bra74} basing on the theory of Teutsch and Hughes \cite{Teu56} developed originally for gases.
The theory successfully describes the growth of Ps yield in strong fields because of the "heating" of the
positron and keeping it in the Ore gap.  Characteristic field strength $E_*$ (when the Ps yield begins to
grow up) may be estimated from the condition ${\rm LET}=eE_*$, where LET (the linear energy transfer of
e$^+$, $-dW/dx$) is determined by excitations of molecular vibrations (typical value is about $10^{-2}$
eV/\AA). Thus we find that the corresponding field is $E_* \sim 1$ MV/cm.

In small fields the situation seems to be different. According to the Brandt theory, the decrement of the Ps yield is explained
by the influence of the field on the positron escaping from the Ore gap during its thermalization. This process is considered
as e$^+$ random walks in energy space with the field dependent diffusion coefficient $D_p(E) \sim (eEl)^2/(l/v)$. It is
expected that, when the field is applied, the positrons should abandon the Ore gap faster and therefore Ps formation must
decrease. However, it is unclear why the authors considered only positron diffusion out of the Ore gap but did not include
their diffusion back to this energy region.  This theory completely ignores the track effects on Ps formation. It is unable to
interpret the suppression of the Ps yield to almost zero when a quasifree electron scavenger is added to the medium. These
effects are beyond the Ore model and the theory \cite{Bra68,Bra74}. Moreover, it predicts too large thermalization lengths for
e$^+$ and e$^-$, 100 times higher than ones known in radiation chemistry.\cite{Mog75} As we shall see in the next sections
these effects in low fields get a natural explanation within the framework of the recombination mechanism of the Ps formation.

In the present paper we have developed new approach for consideration of the electric
field effect on Ps formation at low electric fields ($\lesssim 100$ kV/cm), where it is
possible to neglect positron acceleration by the field up to the eV energies. This
approach is based on the recombination mechanism (diffusion-recombination model or
blob model), \cite{Bya93,Bya96,Ste00} and properly takes into account the multipair
nature of the end part of the positron track.

In Section II Ps formation in condensed molecular media is considered. We introduce a concept of
the quasifree positronium (a weakly bound e$^+$-e$^-$-pair, a precursor of the Ps in a bubble).
We also present the arguments in favor of that the Ore process plays negligible role at low
electric fields, where an adequate description of Ps formation may be achieved in the framework
of the recombination mechanism. Finally the difference between spur and blob models is discussed.
In Section III an approximate mathematical formulation of the problem is given. General
expression for the Ps formation probability is obtained and some particular cases are
considered. The last section contains some discussions of the results and further extension of
this model.

\section{Positronium formation in condensed media}

\subsection{Quasifree Ps state. Modification of the Ore gap}

Consideration of the Ps formation in condensed molecular media (dielectric liquids, polymers, some molecular
crystals) requires clarification of all intermediate stages and states, preceding the formation of the final
equilibrium state of the Ps atom. It is well known that in liquids the repulsive interaction resulting from
the exchange of the electrons between Ps and surrounding molecules eventually leads to the formation of an
equilibrium bubble with the radius about some angstroms. \cite{Nak88} However, as we shall see below,
formation of the quasifree Ps (qf-Ps) precedes the formation of the Ps bubble. The qf-Ps state corresponds
to the bottom of the lower energy band available to the interacting e$^+$-e$^-$ pair before any rearrangement
of molecules takes place. The notion of the qf-Ps was invoked in the frameworks of the recombination
mechanism of the Ps formation in order to explain the E-field effect on Ps formation in hydrocarbons
\cite{Ste00} and in relation to the Ps bubble model.

An availability of a large free space in gases always allows neglecting the zero-point kinetic energy of Ps,
caused by the presence of the gas molecules. So in this case Ps binding energy is simply $-Ry/2$. In
condensed phase the presence of molecules, firstly, essentially increases Ps zero-point kinetic energy
arising owing to Ps repulsion from them (both e$^-$ and e$^+$ are repelled from the cores of atoms because of
exchange and Coulombic repulsions, respectively). Sometimes it is called as a "confinement" of the Ps.
Secondly, molecular electrons screen the e$^+$-e$^-$ Coulombic attraction and increase average e$^+$-e$^-$
separation in the positronium. Obviously these factors decrease binding energy between the positron and
electron constituing the positronium and therefore reduce the width of the Ore gap, which may even completely
disappear.

For the first time the confinement of the Ps was studied by Brandt \cite{Bra60} on the base of the solution found by Sommerfeld
and Welker \cite{Som38,Gro46} of the quantum-mechanical problem of a hydrogen atom confined by an impenetrable sphere of the
radius $R$. Strict analytical solution of the same problem for the Ps atom is not available because of the impossibility of
formulating the problem in terms of the center-of-mass coordinate and relative coordinate when the external potential is
present.

Elementary estimation of the Ps confinement effect is the following. Ground state energy of a
point particle with the mass $2m$ in an infinite potential well of the radius $R$ is
$\frac{\pi^2\hbar^2}{4mR^2} = \frac{Ry}{2}\left(\frac{\pi a_B}{R}\right)^2$, where $a_B=0.53$
\AA\ is the Bohr radius. For $R=\pi a_B \approx 1.66$ \AA\ this energy becomes equal to the
binding energy, $-Ry/2$, of the free Ps atom taken with an opposite sign (for H atom it
happens at $R=1.835a_B = 0.97$ \AA).  This gives a hint, that Ps might become unstable with
respect to a break up on e$^+$ and e$^-$ and indicates that in a dense medium the width of
the Ore gap becomes narrower and even may disappear.

Certainly such an approach simulating Ps repulsion from molecules by an infinitely deep potential
wall is crude. A more realistic evaluation of the energy of qf-Ps in a liquid phase can be
obtained basing on the modified Ps bubble model, \cite{Bya00,Ste00PC} particularly using the
values of the depth $U$ of potential wells of equilibrium Ps bubbles ($U$ does not depend on the
radius of the Ps bubble).

The matter is that the physical meaning of $U$ coincides with that of the Ps work function $V_0^{\rm Ps}$ (the energy needed
for Ps to enter the liquid without any rearrangement of molecules and stay there in the quasifree state). With the free-volume
radius of the bubble tending to zero, the positronium transfers from the bubble just to the qf-Ps state, which has no
preferential location in a bulk. Note, that $V_0^{\rm Ps}$ is an analog of the work function $V_0^-$ of the quasifree (excess)
electron, $\rm e^-_{qf}$, but there is one important distinction. $V_0^-$ is at the same time the ground state energy of the
$\rm e^-_{qf}$, because the energy of the electron at rest after having been removed from the liquid to infinity is defined to
be zero. \cite{Spr68} Ps work function $V_0^{\rm Ps}$ differs from the qf-Ps ground state energy, $V_0^{\rm Ps}-Ry/2$, by a
constant shift, because its energy far outside the liquid is not zero, but $-Ry/2$. However, it is convenient definition
because, would we consider dissociation of Ps in vacuum, the energies of the largely separated e$^+$ and e$^-$ become equal to
zero.

Energetics of the qf-Ps state may be illustrated with the help of the Born-Haber cycle:
\begin{center}
\epsfig{file=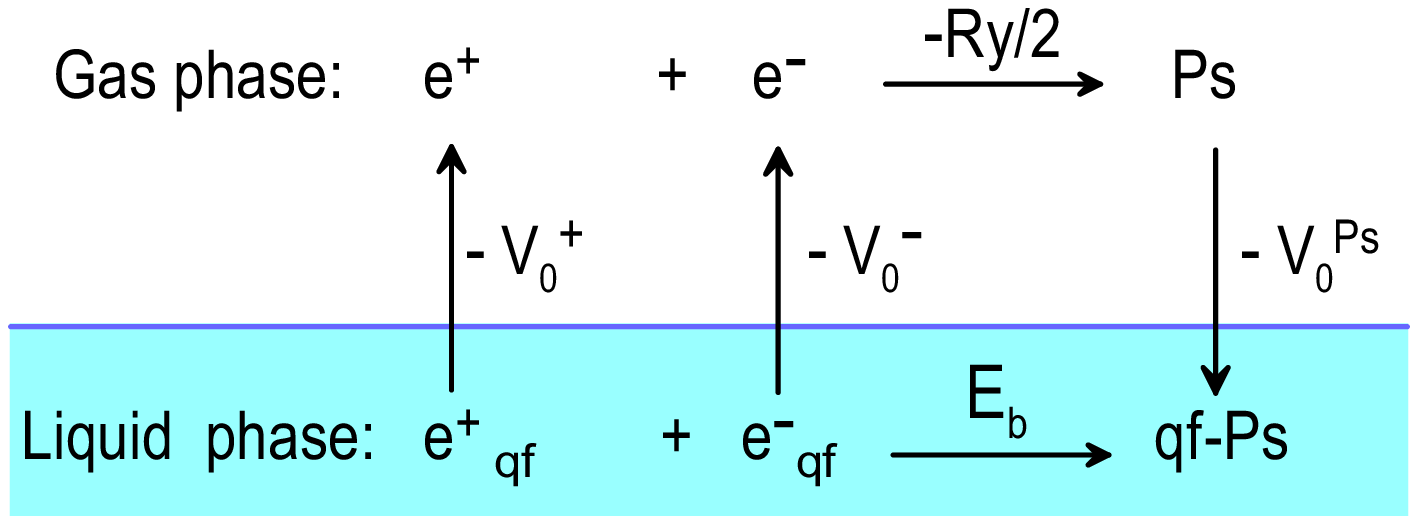, width=100mm}
\end{center}
%
It consists of the following blocks: 1) removal of the quasifree electron and positron from the liquid to
vacuum accompanying with some expenditure (or release) of energy, namely the sum of the respective work
functions, $-V_0^- -V_0^+$, taken with negative sign; 2) formation of the Ps with the binding energy $-Ry/2$
in vacuum; 3) sending back of the Ps atom into the liquid phase supplying it with the energy $V_0^{\rm Ps}$.
From the Born-Haber cycle we are able to estimate binding energy, $E_b$, of the qf-Ps state:
$$
E_b = -V_0^- - V_0^+ + V_0^{\rm Ps} - Ry/2 < 0. \eqno(4a)
$$
It is believed that $V_0^+$ is negative because of prevalence of the polarization interaction of e$^+_{\rm
qf}$ with the medium molecules over repulsion from the nuclei. \cite{Iak82}  So we may roughly equate $V_0^+$
to the energy $P_-$ of the polarization interaction of $\rm e^-_{qf}$, which can be calculated by
decomposition of $V_0^-$ into the sum of kinetic $K$ and polarization $P_-$ terms: $V_0^- =K + P_-$.
\cite{Spr68} For hydrocarbons experimental values of $V_0^-$ are rather close to zero, which means that $K$
and $P_-$ tend to cancel each other.  Using relevant expressions from \cite{Spr68} we obtain that $V_0^+
\approx P_- \approx -(2~~{\rm to}~~3)$ eV. Further, from the experimental data on the ortho-Ps lifetimes and
the widths of the "narrow" component of the 2$\gamma$-angular correlation spectrum accounting curvature
dependence of surface tension, it is possible to obtain $U\equiv V_0^{\rm Ps}$. It was found,
\cite{Bya00,Ste00PC} that practically in all investigated liquids of various chemical nature values of
$V_0^{\rm Ps}$ are close to 3 eV. Thus from Eq.(4a) we obtain that $E_b \approx -(0.5~~{\rm to}~~1)$ eV. It
means that qf-Ps is a loosely bound structure although the long-range Coulombic attraction between e$^+$ and
e$^-$ always provides an existence of the bound state (at least within the Onsager radius). Because
positron-electron separation, $r_{ep}$, in qf-Ps is expected to be large we may rewrite $E_b$ in the
following form:
$$
E_b \approx -{Ry \over 2}\cdot {1\over \varepsilon_\infty^2} \cdot {3 a_B \over r_{ep}}, \eqno(4b)
$$
where $\varepsilon_\infty$ is the high-frequency dielectric permittivity of the liquid, and $3 a_B$ is the distance between
e$^+$ and e$^-$ in the Ps atom in vacuum (Ps diameter is $\langle\varphi|r_{ep}|\varphi\rangle/\langle\varphi|\varphi\rangle =
3a_B$, where $\varphi \propto \exp(-\mu r_{ep}/a_B)$ and $\mu=1/2$ accounts for the reduced mass factor). The scaling $E_b
\propto \varepsilon_\infty^{-2}$ directly follows from the Schr\"odinger equation for Ps and is confirmed experimentally.
\cite{Gul88} Using above mentioned numerical assessment for $E_b$ and Eq.(4b) we conclude that $r_{ep}$ is about 5 \AA. In such
a Ps state delocalized over $r_{ep}$ the density of the "native" electron on the positron is rather small and e$^+$ will
primarily annihilate with electrons of the medium with the average lifetime about 0.5 ns and contribute to the free e$^+$
component of the ACAR spectrum. Thus, qf-Ps component could hardly be resolved experimentally, but it should renormalize
contribution ascribed to the "free" e$^+$ annihilation. If the standard three exponential decomposition of the LT spectra is
used the presence of qf-Ps is revealed, for example, in anomalous behavior of free e$^+$ lifetime. \cite{Gow00}

Further evidence that qf-Ps is a delocalized object is provided by rather high values of
$V_0^{\rm Ps}$. Firstly, if we accept for the moment that qf-Ps in liquid may be considered as a
"point" particle, in that case $V_0^{\rm Ps}$ should represent just its zero-point kinetic
energy due to the repulsion from surrounding molecules. Secondly, it is quite reasonable to
assume that an excess quasifree electron injected into the liquid undergoes the same exchange
repulsion as the "point"-qf-Ps. Therefore, it immediately implies that zero-point kinetic
energy, $K$, of the e$^-_{\rm qf}$ should be twice as large as that of positronium and equal to
$2V_0^{\rm Ps} \approx 5$-7 eV. It is simply because the electron is half as heavy as Ps.
However, it is very difficult to conceive that the negative polarization contribution, $P_-$,
could compensate $2V_0^{\rm Ps}$ term and yield known experimental values of $V_0^-$, which are
usually close to zero or even slightly negative.

Finally we are able to estimate the low boundary $W_{low}$ of the Ore gap in condensed phase. Because the Ore
process is just an electron transfer reaction, we assume that no rearrangement of molecules occurs and,
therefore, final positronium will be in the quasifree state (formation of the bubble requires much longer
time).  If translational kinetic energy of the final qf-Ps is thermal $(\approx 0)$, the excess kinetic
energy, $W$, of the projectile positron in this case should coincide with the low boundary the Ore gap: $W =
W_{low}$. Now, let us carry out the Born-Haber cycle again, i.e. virtually pull out reagents to the gas
phase, accomplish Ps formation reaction (1) and return products to the liquid:
\begin{center}
\epsfig{file=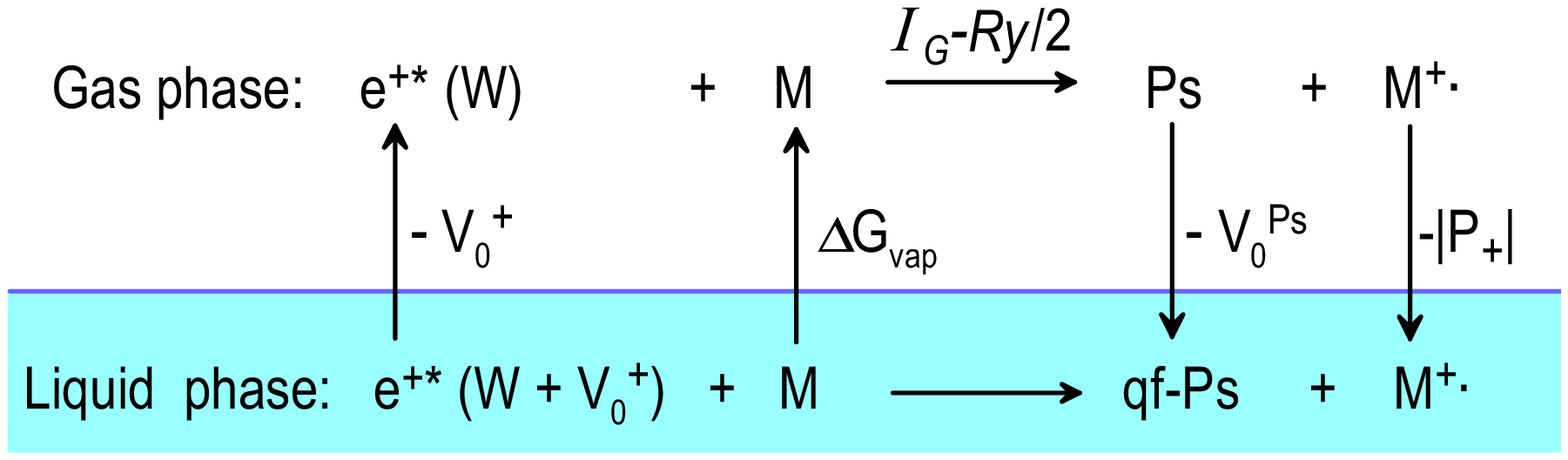, width=120mm}
\end{center}
In parenthesis we displayed the total energy of the e$^+$ in the liquid and gas phases. $\Delta G_{vap}$ is
the vaporization energy of the molecule (a few tenth of eV) and $P_+$ (negative) is the energy of
polarization interaction of the radical-cation M$^{+\cdot}$ with environment. Calculating total energy
balance, one obtains that Ps formation may proceed if $W+V_0^+ - \Delta G_{vap} -(I_G-Ry/2) -V_0^{\rm Ps}
+|P_+| >0$.  Using an approximate relationship between the liquid phase and the gas phase ionization
potentials, $I_L = I_G - |P_+| + V_0^-$, it is possible to write the following relationship for $W_{low}$:

$$
W_{low} = I_L - V_0^- - V_0^+ + \Delta G_{vap} + V_0^{\rm Ps} - Ry/2.
\eqno(5)
$$
Adopting that $I_L -V_0^- \approx 9$ eV (Table I) and taking into account previous estimations
for other quantities involved, we obtain that $W_{low}$ is 7-8 eV.  It is close enough (or may
be even higher) to the threshold $W_{ex}$ of electronic excitations.  Thus, in molecular media
the Ore gap $W_{ex}-W_{low}$ is rather narrow or may completely disappear, $W_{ex} < W_{low}$.
Smallness of $W_{ex}-W_{low}$ together with rather high value of the e$^+$ energy losses
($\sim 10^{-2}$ eV/\AA\ for subionizing particle) \cite{Ste95,Bya96} related mainly to
excitation of molecular vibrations, make e$^+$ residence time inside the Ore gap very short
(less and probably much less than 0.1 ps).  That is why, generally speaking, Ps formation
according to the Ore mechanism is ineffective in molecular media.  However, it appears to be
important when high external electric fields are applied.

\subsection{Recombination mechanism of Ps formation. Spur and blob models}

Recombination mechanism postulates that Ps formation proceeds via combination of the positron
and one of the knocked out electrons in the terminal part of the positron track.  There are two
models, which utilize this mechanism, namely, the spur model \cite{Mog95,Mog74} and the blob
model (or diffusion-recombination model) \cite{Bya85,Bya74,Bya76,Bya93}. According to these
models the decrease of the Ps yield in low fields (up to 100 kV/cm) occurs because the field
takes apart the positron and track electrons in different directions. Therefore, probability of
their encounter, as well as the Ps yield, drop down. In spite of the fact that both models
answer the question about the Ps precursor in the same way, they differ in the point what is the
terminal part of the e$^+$ track and how to calculate the probability of the Ps formation.
Quantitative formulation of the spur model was given by Tao. \cite{Tao76} It is based on the
following assumptions:

1) the positron and secondary electrons knocked out during of the last several acts of
ionizations thermalize within the same volume of the medium. Positive ions also reside in the
same space;

2) secondary electrons recombine with the same probability either with the positive ion or the
positron. Therefore, the probability that the positron gets an electron for subsequent Ps
formation is equal to $n_0/(1+n_0)$, where $n_0$ is the "initial" number of ion-electron pairs
in the terminal spur;

3) probability of the "elementary act" of e$^+$-e$^-$ recombination is taken in the Onsager
form, $1-{\rm e}^{- r_c/b}$, where $r_c = e^2/\varepsilon T$ is the Onsager radius, $T$ is the
temperature in energy units, $b$ is the "initial" distance between e$^+$ and e$^-$ by the end
of their thermalization, and $\varepsilon$ is the dielectric permittivity of the medium.

Thus, in the frameworks of the spur model Ps formation probability is written as:

$$
P_{\rm Ps} = \frac{n_0}{1+n_0} (1-{\rm e}^{-r_c/b}) \cdot
             {\rm e}^{-\lambda_f t_{\rm Ps}}.
\eqno (6)
$$
The last exponential factor takes into account a possibility of the free positron annihilation
during the  Ps formation time $t_{\rm Ps}$ (some picoseconds) with the annihilation rate
$\lambda _ f $ ($\lesssim 2$ ns$^{-1}$). Obviously, contribution of this factor is negligible.

Because in all nonpolar molecular media $ \varepsilon \approx 2$ at room temperature $r_c$ is
equal to $\approx 300$ \AA. On the other hand, radiation chemistry data \cite{Fri87} show that
typical thermalization lengths $b$ of electrons are $\lesssim 100$ \AA. Thus, the Onsager
factor, $1-{\rm e}^{-r_c/b}$, is also very close to unity. Therefore, to explain observable
values of Ps yields, which never exceed 0.7, we have to adopt in accordance with Eq.(6) that
the terminal positron spur contains in average 2-3 ion-electron pairs.

Later on, the recombination mechanism of Ps formation became extremely widespread.  It is able to interpret extensive data on
Ps chemistry (see \cite{Mog95} and ref.  therein). In contrast to the Ore model, this mechanism easily explains variation of
the Ps yields from 0 up to 0.7 in the substances of various chemical nature, where the Ore gap is practically the same. Changes
of the Ps formation probability under phase transitions also received natural explanation. Experimentally observable monotonic
decrement of the inhibition curves of the Ps yields (practically down to zero) in solutions of electron acceptors contradicts
the Ore model, but is well incorporated to the recombination mechanism.  It naturally explains antiinhibition and
antirecombination effects, including experiments on formation Ps in low electric fields in pure liquids and mixtures.
\cite{Wan98,Ito79,Ito88}

In \cite{Bil98} the spur model was used as the basis for computer simulation of the Ps formation process.
Terminal part of the e$^+$ track was approximated by a linear sequence of separated ion-electron pairs.
Fitting of the data in 2,2-dimethylbutane, it was obtained that the average distance between final acts of
ionizations at the end of the e$^+$ track is about 165 \AA, which seemingly agrees with the spur model.
Nevertheless, having considered the data for more than 50 liquids, the authors \cite{Bil98} came to the
conclusion that multiparticle effects in the terminal positron spur are rather important for correct
description of the processes in it.

However, inconsistencies of the spur model should be mentioned (and the conclusions of the above simulations, in particular)
with the data on the structure of the positron track (Fig.\ref{e_track}). The matter is that the end of the positron track is
not a spur, containing 2-3 ion-electron pairs, but the blob, containing about 30 overlapped ion-electron pairs.
\cite{Bya96,HRC91,Ito96,Ito97} In essence this statement relies on the following two well known facts related to behavior of
the energized electron (positron) on the blob formation stage, i.e. at energies $Ry\lesssim W\lesssim 500$ eV (Appendix A):

1) Ionization energy losses of such a positron (electron) is about 1-4 eV/\AA. It means, that
e$^+$ very effectively ionizes molecules at the end of its track. Because the average energy
$W_{iep}$, needed for creation of one ion-electron pair is \cite{Bya85}~ $1/G_{iep} \approx
16$-22 eV, ~($G_{iep}$ is the initial yield of ion-electron pairs per 100 eV of the absorbed
energy) the distance $l_i$ between subsequent ionizations made by e$^{+*}$ is about
$W_{iep}/{\rm LET} \approx 10$-20 \AA. Dependence of $l_i$ vs e$^+$ energy is shown in
Fig.\ref{a_bl_W_bl};

2) Motion of the positron in the blob is diffusion-like (its transport path at these energies varies from 20
down to several \AA, Fig.\ref{a_bl_W_bl}). Thus, efficient e$^+$ scattering at the blob formation stage
results in approximately spherical shape of the blob. Therefore, the simulation of the end part of the e$^+$
track by the straight-line sequence of isolated ion-electron pairs \cite{Bil98} or by a cylindrical column of
ionizations \cite{Ito97} seem to be unjustified.

The statement that the terminal part of the e$^+$ track is a blob, but not a spur is not just a question of
terminology. As we shall see below, some primary processes, including Ps formation, proceed in a different
way in blobs and spurs.\cite{Ste00} Another important distinction is that in the blob model behavior of the
positron is considered as very different from that of intrablob electrons and ions. Usually in the blob
approach it is assumed that e$^+$ is rather mobile and easily escapes from its blob (Sec.II E). Particularly,
it implies that the multiparticle nature of the terminal part of the e$^+$ track cannot be correctly taken
into account via the factor $n_0/(n_0+1)$. In details we shall discuss these differences below, but now we
shall briefly remind the life history of the positron, starting from its birth up to the annihilation.

\subsection{Life history of the positron}

Even though the positron is a stable elementary particle, its lifetime in matter turns out to be
very short owing to presence its antiparticles -- electrons. Usually the positrons are born in
nuclear $\beta^+$-decay (of $^{22}$Na or $^{64}$Cu) with typical initial energy about several
hundreds of keV. After implantation into a medium their life may be divided into three stages.
It is convenient to subdivide the first stage into two steps: a) slowing down by ionization and
b) thermalization. Retardation from the initial energy down to the ionization threshold lasts
about $10^{-11}$ s.  When e$^+$ energy becomes less than $W_{bl}\sim 500$ eV, the positron forms
its terminal blob (Appendix A). In the second step, the subionizing positron and knocked-out
electrons become thermalized, exciting primarily intra- and intermolecular vibrations.  In
molecular liquids, it usually takes less than $10^{-13}$ s in addition to the duration of the
first step.

The second stage involves fast intrablob reactions with the participation of primary
radical-cations RH$^{+\cdot}$ (ions) and thermalized electrons and positrons. According to the
recombination mechanism, Ps is formed just at this stage.  This stage is limited by
localization (solvation), ion-electron recombination, or out-diffusion of the species. The
typical duration of this stage is $10^{-12}$-$10^{-11}$ s.

The third stage includes different chemical reactions between Ps, localized species, and
additives. It is terminated by e$^+$ annihilation from para-Ps, "free"-e$^+$, or ortho-Ps by
the pick-off process. This final stage lasts about $10^{-10}$-$10^{-9}$ s.

\subsection{Main distinctions between spurs and blobs}

\subsubsection{Sizes and energy release}

In radiation chemistry the term "spur" denotes the small isolated spherical part of the track of an ionizing particle,
containing a few ($n_0^{sp}=1$-5) ion-electron pairs. \cite{HRC91,Par87} Average energy release in the spur is 30-50 eV.  Its
typical size is determined by a thermalization length of the knocked out electron in the Coulombic field of the ion. In a crude
description of radiation-chemical processes it is possible to assume, that tracks of high energy electrons and Compton
electrons produced by $^{60}$Co $\gamma$-rays consist only of isolated spurs (contribution of short tracks and blobs is rather
small). In spurs with a few ion-electron pairs, Onsager's description of the geminate recombination is plausible. \cite{Hum74}

As we already mentioned, the end part of the e$^+$ track is a blob. Calculations show (Appendix A) that the spatial
distribution of ionization events in it can be simulated by gaussian function
$n_0\frac{\exp(-r^2/a_{bl}^2)}{\pi^{3/2}a^3_{bl}}$. Here $n_0=W_{bl}/W_{iep}\approx 30$ is the initial number of ion-electron
pairs and $a_{bl} \approx 40$ \AA. The blob size $a_{bl}$ is determined by the ionization slowing down of the energetic
positron, when it loses energy from $W_{bl} \sim 500$ eV (total energy deposition in the blob) down to the ionization
threshold.

\subsubsection{Quasi-neutrality condition}

In blobs knocked out electrons before getting localized rapidly adjust their motion to the
distribution of the primary positive ions and efficiently screen them \cite{Pai00}. Such a
quasi-neutrality electron distribution in the field of ions is established because 1) Coulombic
interaction between ions and electrons is strong enough (the Onsager radius is much larger than
the blob size and average electron thermalization displacement\cite{Fri87})~ 2) mobility of
electrons is much higher than that of ions. Simple estimations (Appendix B) show that the width
of the spatial distribution of the intrablob electrons is only a few angstroms larger than that
of ions: $a_e - a_i \approx 2.4 a^2_{bl}/r_c n_0 \ll a_{bl}$. Hence $a_{bl} \approx a_e \approx
a_i$. Here $a_e^2$ and $a_i^2$ are dispersions of the distributions of the electrons and ions,
respectively.  The concept of quasi-neutrality of the blob seems reasonable especially in
non-polar liquids where an absence of significant spatial fluctuations of local electric field
(in contrast to polar media) favors establishing blob quasi-neutrality and complete recombination
of charges.

\subsubsection{Expansion of the blob}

Because of the Coulombic attraction between charged intrablob species their diffusion expansion does not
proceed independently.  As a result out-diffusion of electrons is almost completely suppressed, but the
diffusion coefficient of ions is increased by a factor of 2. This case is known as ambipolar diffusion
(Appendix C). Thus blob expansion proceeds very slowly and may be practically neglected in the problem of Ps
formation.

\subsubsection{Elongated thermalization and recombination}

After the last ionization event in the terminal blob, the positron and intrablob electrons become subionizing. Having no
possibility to excite electronic transitions in molecules, they lose energy by exciting molecular vibrations. In polar media
there is an additional way of energy losses: reorientation of the dipole molecules (Debye losses).  Thermalization theories
usually consider molecular excitation a result of its perturbation by an alternating electric field induced by a flight of an
isolated charged particle. \cite{Tac87,Ste95} If the Fourier spectrum of this field contains harmonics with frequencies close
to those of molecular vibrations, excitations get very probable. In the blob because of the presence of a large number of
ion-electron pairs and efficient screening (local quasi-neutrality), the resulting electric field at a given point does not
fluctuate significantly. Thus, in spite of fast motion of intrablob electrons, they cannot lose kinetic energy easily via
excitation of vibrations. This lengthens thermalization of electrons and ion-electron recombination because these processes
depend on the absorption rate of released energy.

\subsection {Behavior of the positron and positronium formation}

\subsubsection{e$^+$ escape from the blob}

Because the blob is electrically neutral, Coulombic interaction between the positron and its
blob is absent. Polarization attraction between e$^+$ and the blob is small enough (Sec.V F).
Therefore, after the last ionization even subionizing positron may easily to escape from its
blob during thermalization by travelling a longer distance than the blob electrons, which are
held by the ions.  So one may expect that the width of the e$^+$ spatial distribution by the end
of thermalization is larger than the blob size:  $a_p > a_{bl}$.  Strictly speaking $a_p^2$ is a
dispersion of initial (gaussian) distribution of the thermalized positron around the center of
the blob.  Because e$^+$ thermalizes primarily outside the blob its thermalization time is not
affected by the above mentioned elongation of the thermalization process.

\subsubsection{e$^+$ mobility}

In non-polar liquids the positron after its thermalization is expected to be in the quasifree
state having rather high mobility in comparison with than that of the quasifree electron (Table
I). Firstly, it is due to the absence significant spatial fluctuations of local electric field,
which may trap or localize the positron (the medium is nonpolar). Secondly, a density of a
liquid state is too high to permit clasterization of molecules around the positron like in
pressurized gases. \cite{Iak82}

Actually because of the screening, Coulombic interaction with nuclei plays minor role in the
scattering of the quasifree positron and electron. Their mobilities are primarily determined by
the small angular scattering on outer molecular electrons (see, for example, Eq.12 in
ref.\cite{Bya96}). However, for excess electrons there is some spin-exchange contribution which
enhances e$^-$ scattering and decreases their mobility in comparison of the e$^+_{\rm qf}$ one.
Exchange repulsion may even lead to full localization of the excess electron in a void.
Contrary, the positron prefers to reside in a bulk because of polarization interaction.
Preferential e$^+$ traps are positive density fluctuations (aggregation of
molecules).\cite{Iak82} However, such traps practically cannot be formed in a liquid phase
because of strong asymmetry of intermolecular interaction potential.

\subsubsection{"Elementary act" of Ps formation}

Being thermalized primarily outside its terminal blob, quasifree positron may diffuse to the
blob again and pick up one of the intrablob electrons. That process needs the "initial"
separation between the particles not exceeding the average distance between intrablob species,
i.e.  $(4 \pi a_{bl}^3/3n_0)^{1/3} \approx 20$ \AA. According to Eq.(4b) initial binding energy
of this e$^+ \cdots$ e$^-$ pair is about 0.1 eV. Translational kinetic energies of each particle
composing the pair must be less than the initial binding energy, otherwise the pair will break
up. Therefore e$^+$ and e$^-$ just before the "elementary act" of Ps formation must be practically
thermalized.

In the framework of the adiabatic approximation (the positions of molecules are fixed) such a pair is not at the bottom of its
energy spectrum and may continue to release energy via excitation of intra- and intermolecular vibrations and finally form at
equilibrium (in adiabatic sense) the qf-Ps state discussed in Sec.II.A. In comparison with the binding energy of qf-Ps, the
e$^+ \cdots$ e$^-$ pair possesses about 1 eV of excess energy, which is accumulated in the form of a potential energy of the
mutual Coulombic attraction. Let us now see how this pair transforms into qf-Ps and what happens with this excess energy during
e$^+ \cdots$ e$^- \to $ qf-Ps transformation.

Let total momentum of the e$^+ \cdots$ e$^-$ pair be zero at the beginning. When particles getting closer, kinetic energy of
both the positron and electron increases. If then one of the particles creates a phonon (excites a molecular vibration) its
energy drops down (as well as momentum and it changes direction also) and as a result the total momentum of the e$^+ \cdots$
e$^-$ pair becomes non zero (just due to momentum conservation in inelastic phonon scattering). This mechanism gives rise to
the motion of the e$^+ \cdots$ e$^-$ pair. Adopting $\approx 0.01$ eV/\AA\ as a typical value of the e$^-$ (e$^+$) LET  at
energies about 1 eV due to the excitation of vibrations, we may estimate total path of the e$^+ \cdots$ e$^-$ pair before it
becomes qf-Ps. It is $E_b/{\rm LET} \approx 50$ \AA\ (the value of $E_b$ is estimated in Eq.(4a)). When the positron and
electron get closer, their polarization interaction with the medium decreases. Larger number of surrounding molecules become to
"feel" e$^+ \cdots$ e$^-$ pair as an electrically neutral object. So LET of such a pair may decrease and its total path may
increase.

After losing all its excess potential energy the e$^+ \cdots$ e$^-$ pair becomes the qf-Ps, which is in
thermal equilibrium with the environment. In a disordered medium (for example, in a liquid) it immediately
finds nearest place with slightly lower density and will stay there, if the temperature cannot push it out
from that shallow trap. This state may survive for a rather long time (it may be comparable with the free
e$^+$ lifetime) in a solid medium, where molecular rearrangements are energetically forbidden or require
longer time.

In liquids (or "soft" media) qf-Ps state continuously evolves into the bubble state at a longer
time scale because of further gain of the e$^+$-e$^-$ Coulombic attraction and decrease of the
positronium zero-point kinetic energy. It becomes possible due to rearrangement of molecules and
appearance of additional free space. This is the beginning of the Ps bubble formation stage.
Kinetic energy released during compactification of the Ps stimulates further local decrease of
the density.  Eventually an equilibrium bubble with the Ps inside is formed.  Schematically Ps
formation may be displayed as follows:
$$
\rm e^+_{qf} + e^-_{blob} ~\to e^+ \cdots e^- ~\to ~\hbox{qf-Ps} ~\to ~\hbox{Ps
in the bubble}.
\eqno(7)
$$

Cursory consideration of the bubble growth based on the solution of the Euler equation for
incompressible fluid taking into account the Laplace pressure and the quantum-mechanical
"pressure" exerted by the Ps atom on the wall of the bubble, gives the following estimation for
the bubble formation time
$$
t_{bub} \approx 3 \hbox{ ps } \cdot \sqrt{\rho \hbox{, g/cm$^3$}}
    \left( \frac{25}{\sigma \hbox{, din/cm}} \right)^{7/8},
$$
where $\rho$ is the density and $\sigma$ is the surface tension coefficient of the liquid. It implies that
bubble formation lasts about several picosecond in the majority of liquids. On the similar basis
consideration of the self-localization of the positronium (the last step of Eq.(7)) was suggested by Khrapak.
\cite{Kra95}

Perhaps molecular crystals (anthracene, acenaphthylene) give us the best example illustrating the
transformation of qf-Ps into the bubble state. It is known that in these substances in the solid
state there are neither longlived nor "narrow" components in the lifetime and  angular
correlation (ACAR) spectra, respectively. So from a viewpoint of experimentalists the Ps atom
does not form there. However when the temperature is slightly above the melting point (in the
liquid phase) Ps formation probability becomes very large (more than 0.6). \cite{Gow76,Gow78}
According to our model, it is due to impedance in the solid phase of the qf-Ps transformation to
the bubble state, but in the liquid phase it easily proceeds. As a result, in the solid phase we
expect contributions both from free e$^+$ and qf-Ps annihilations, but it might be difficult to
resolve them experimentally. Annihilation parameters of qf-Ps (lifetime and width of its ACAR
component) may be obtained by subtracting from the solid-state annihilation spectra the free
e$^+$ contribution, which must be taken (with the same intensity and lifetime; corrections
accounting for the variation in the average electron density at the phase transition may be
needed) from the measurements in the liquid phase.

\subsubsection{Role of localized species}

From a viewpoint of the energy balance localized electron and positron might participate in Ps formation. Actually $\rm
e^-_{loc}$ already resides in a small void and its binding energy there is about 1 eV. Hence, Ps formation may proceed through
$\rm e^+_{qf}$ trapping by $\rm e^-_{loc}$ on an appropriate Rydberg-type orbit with a subsequent formation of the equilibrium
Ps bubble.

However, in non-polar liquids, as we shall see below, electron localization takes much longer than Ps
formation and, therefore, $\rm e^-_{loc}$ does not exist on a time scale of Ps formation process.

In polar liquids, like water, one may expect that e$^+$ gets also localized within a time comparable with electron
localization.  After that, mobility of the particles drastically drops down and they simply have not enough time to meet each
other (time is limited by a free positron lifetime $\sim 0.5$ ns). Really the diffusion displacement of $\rm e^+_{loc}$ during
0.5 ns is smaller than thermalization displacement ($\sim a_p$).

Thus, in what follows we do not include localized particles into equations describing Ps
formation.

\section{Early intrablob processes: general formulation}

Thus, according to the blob model we assume that Ps formation proceeds via recombination of the
thermalized but not yet localized positron with one of the intrablob quasifree electrons,
released while the positron is being slowed down by ionization:

$$
{\rm e^+_{qf} + e^-} \stackrel{k_{ep}}{\longrightarrow} \hbox{qf-Ps}
    \longrightarrow \hbox{Ps in the bubble}.
\eqno(8a)
$$
This reaction may compete with other intratrack processes

$$
{\rm localization:  \qquad \qquad\qquad \qquad \qquad \qquad
    e^-} \stackrel{\tau^{-1}_e}{\longrightarrow}  {\rm e^-_{loc},
    \qquad
    e^+_{qf}} \stackrel{\tau^{-1}_p}{\longrightarrow} \rm e^+_{loc},
\eqno(8b)
$$
$$
{\rm \hbox{ion-electron recombination (IER):} \qquad
    e^- + RH^{+\cdot}} \stackrel{k_{ie}}{\longrightarrow}
               \rm RH^*  \to \hbox{Products}.
\eqno(8c)
$$
$$
\rm \hbox{ion-molecule reaction:}  \qquad\qquad  \qquad \qquad \qquad
    RH^{+\cdot} + RH \to R^\cdot + RH^+_2,
\eqno(8d)
$$

In C$_6$F$_6$ intratrack electrons by the end of their thermalization can be weakly captured by host molecules:
\cite{Gan76,End82,Ito79}

$$
\rm e^- + C_6F_6 \to C_6F_6^-.
\eqno(9a)
$$
However, the positron may easily pick up such an electron from C$_6$F$_6^-$ ion and form Ps:

$$
\rm e^+_{qf} + C_6F_6^- \to Ps + C_6F_6.
\eqno(9b)
$$
Therefore in this case $c_e({\bf r},t)$ in Eqs.(10) will stand not for the concentration of the electrons (as in other
liquids), but for the concentration of $\rm C_6F_6^-$ anions.

The blob model describes nonhomogeneous kinetics of reactions (8) in terms of concentrations of the particles, because the
number of particles involved is large and local motion of quasifree electrons and the positron is fast enough in comparison
with Ps formation time (Sect.V E). In the simplest case the external electric field enters diffusion-recombination equations
(10) via a drift of the positron distribution (electrons are tightly bounded to ions):

$$
{\partial c_i({\bf r},t) \over \partial t} =
   D_{amb} \Delta c_i - k_{ie} c_i c_e - c_i/\tau_i,
\eqno(10a)
$$
$$
{\partial c_e({\bf r},t) \over \partial t} =
     D_{amb} \Delta c_e - k_{ie} c_i c_e - k_{ep} c_e c_p -
                           c_e/\tau_e,
\eqno(10b)
$$
$$
{\partial c_p({\bf r},t) \over \partial t} =
   D_p \Delta c_p - \hbox{div}({\textbf{\textit{v}}}_p c_p)
    - k_{ep} c_e c_p - c_p/\tau_p.
\eqno(10c)
$$
Here $c_i({\bf r},t)$, $c_e({\bf r},t)$, $c_p({\bf r},t)$ are the concentrations of the primary radical-cations RH$^{+\cdot}$
(ions), intrablob electrons and the positron, respectively, at time $t$ at point {\bf r} (measured from the center of the
blob),  $\Delta$ is the radial part of the Laplace operator, $D_{amb} = 2D_i$ is the coefficient of ambipolar diffusion
(Appendix C), $D_i$ and $D_p$ denote the diffusion coefficients of ions ($i$) and the positron ($p$), respectively, and
$k_{ie}$ and $k_{ep}$ are the recombination rate constants.  Decay rate $\tau^{-1}_e$ includes electron localization as well as
possible capture by host molecules, accompanied by formation of their radical-anions (in this work we study pure liquids only).
Similarly, $\tau^{- 1}_i$ is the rate of the ion-molecule reaction and $1/\tau_p=1/\tau_2+1/\tau_p^{loc}$ accounts for the
e$^+$ annihilation ($\tau_2$ is the free positrons lifetime, Table I) and possible localization of the positrons with a
characteristic time $\tau_p^{loc}$.

As discussed above and in Appendix B, blob species distribution is only slightly disturbed by the application of an external
electric field, while the behavior of the positron is very sensitive to the presence of the field.  The field decreases Ps
formation by removal of the positron away from the blob, which suppresses reaction ($8a$). In Eqs.(10) this effect is
approximately taken into account through the term, div(${\textbf{\textit{v}}}_p c_p)$, describing the drift of the positron in
external field {\bf D}. However, we skip polarization interaction between the blob and the positron (see Sec. V G) and neglect
deviation of the field from its average value within the blob caused by the presence of ion-electron pairs. We also assume
linear proportionality between the drift velocity ${\textbf{\textit{v}}}_p$ and {\bf D}, i.e., ${\textbf{\textit{v}}}_p = b_p
{\bf E} = b_p {\bf D}/\varepsilon$ ~($b_p= e D_p / T$ is the positron mobility and $\varepsilon$ is the static dielectric
permittivity of the medium; see comments to Table I). This assumption is justified for $D < 100$ kV/cm where the nonlinear
effect is not significant. A more exact way to account for real field distribution at the terminal positron blob is a joint
solution of Eqs.(10) and an equation on the electric field ${\bf E}({\bf r})$ as was done in Appendix C.

Because the motion of the knocked-out electrons during blob formation and subsequent thermalization obeys the quasi-neutrality
condition, it is natural to adopt the following initial conditions for Eqs.(10):
$$
c_i({\bf r},t\hbox{=0}) =
c_e({\bf r},t\hbox{=0}) =
   n_0 \cdot {\exp(-r^2/a^2_{bl}) \over \pi^{3/2} a^3_{bl}},
    \qquad
c_p({\bf r},t\hbox{=0}) =
 {\exp(-({\bf r}- {\bf l}_p)^2/a^2_p) \over \pi^{3/2} a^3_p}.
\eqno(11)
$$
$n_{0}$ is the initial number of ion-electron pairs and $a_{bl}^2$ and $a_p^2$ are the dispersions of spatial distributions of
the intrablob ion-electron pairs and the positron. These distributions do not take spatial electron-positron correlations of
into account. In contrast with the blob electrons, the motion of subionizing positron is diffusion-like and it may easily
escape from the blob. Hence, dispersion $a_p^2$ of the initial positron distribution may significantly exceed that of intrablob
species. In contrast to $a_{bl}$, $a_p$ is determined by thermalization of the positron, i.e., by its ability to excite intra-
and intermolecular vibrations, while the distribution of the ions in the blob (i.e. $a_{bl}$) depends on a much more efficient
energy loss process -- slowing down of the energetic positron by ionization, when its energy is reduced from $W_{bl}$ to the
ionization potential $I_L$.  In Eq.(11) ${\bf l}_p$ accounts for the drift of the subionizing positron during its
thermalization:
$$
{\bf l}_p = \int_0^{t_{th}} {\textbf{\textit{v}}}_p(t) dt =
    \frac{e{\bf D}}{\varepsilon T} \int_0^{t_{th}}
    D_p(t) dt = \frac{e{\bf D} (a_p^2-
    a_{bl}^2)}{6\varepsilon T}.
\eqno(12)
$$
Here we used that the average e$^+$ thermalization displacement squared $6\int_0^{t_{th}} D_p(t)
dt$ during e$^+$ thermalization time $t_{th}$ is equal to $a_p^2-a_{bl}^2$.

To calculate Ps formation probability, $P_{\rm Ps}$, we must integrate the term $k_{ep} c_e c_p $ over the whole space and
time:
$$
P_{\rm Ps} =
    k_{ep} \int_0^\infty dt \int c_e({\bf r},t) c_p({\bf r},t)d^{\sl 3}r.
\eqno(13)
$$
This is the probability of formation of qf-Ps. However in liquids we expect that this state transforms to the bubble state
rather fast, within several picoseconds as was discussed in the previous section.

Within the framework of the prescribed diffusion method \cite{Fra67} the solutions $c_j({\bf r},t)$ can be written in the
following form:
$$
c_j({\bf r},t) =n_j(t) G_j({\bf r},t),
    \qquad
\int G_j({\bf r},t) d^{\sl 3}r = 1,
\qquad
j = \{i,e,p\}.
\eqno(14)
$$
$n_e(t)$ and $n_i(t)$ are the numbers of quasifree electrons and ions survived up to time $t$, and $n_p(t)$ is the free
positron survival probability. $G_j({\bf r},t)$ is the Green function of a simple diffusion equation (without recombination and
decay terms):
$$
G_j({\bf r},t) =
{\exp[-({\bf r}- {\bf l}_j-{\textbf{\textit{v}}}_j t)^2/
                (4{\cal D}_j t+a_j^2)]
                 \over
                   [ \pi ( 4{\cal D}_j t + a^2_j) ]^{3/2}}.
\eqno(15)
$$
We introduced different subscripts for ${\textbf{\textit{v}}}_j$, ${\cal D}_j$, $a_j$ and ${\bf l}_j$ simply to maintain
symmetry. In accordance with Eqs.(10) and Appendix C ${\textbf{\textit{v}}}_i \to 0$, ${\textbf{\textit{v}}}_e \to 0$, ${\cal
D}_i \equiv {\cal D}_e \equiv D_{amb} \ll {\cal D}_p \equiv D_p$, $a_i \approx a_e \approx a_{bl}$ and ${\bf l}_i={\bf l}_e=0$.
Substituting Eq.(15) and (14) into Eqs.(10) and integrating the resulting equations over whole space, we obtain a much simpler
system of ordinary differential equations on $n_j(t)$:
$$
\dot n_i = - k_{ie} n_e n_i/\tilde V_{ie} - n_i/\tau_i, \qquad
n_i(0) = n_0, \eqno(16a)
$$
$$
\dot n_e = - k_{ie} n_e n_i/\tilde V_{ie}
           - k_{ep} n_e n_p/\tilde V_{ep} - n_e/\tau_e, \qquad
n_e(0) = n_0, \eqno(16b)
$$
$$
\dot n_p = - k_{ep} n_e n_p/\tilde V_{ep} - n_p/\tau_p, \qquad
n_p(0) = 1, \eqno(16c)
$$
where

$$
\frac{1}{\tilde V_{jk}(t)} = \int G_j G_k d^{\sl 3} r =
    \frac{1}{V^0_{jk} (1+t/\tau_{jk})^{3/2}}
    \exp \left[ -\frac{ ({\bf l}_j + {\textbf{\textit{v}}}_j t -
                           {\bf l}_k - {\textbf{\textit{v}}}_k
                        t )^2} {(a_j^2+a_k^2)(1+t/\tau_{jk})}
                        \right]
\eqno(17)
$$
$$
V^0_{jk} = [\pi (a^2_j+a^2_k)]^{3/2},
\qquad
\tau_{jk} = {(a^2_j+a^2_k) \over 4({\cal D}_j+{\cal D}_k)},
\qquad
j,k = \{i,e,p\}.
$$

In Eq.(16$b$) for $n_e$, it is possible to omit the term $k_{ep} n_e n_p/ \tilde V_{ep}$,
because it has a negligible effect on the disappearance of the intratrack electrons. Then within
the prescribed diffusion approach Eq.(13) is rewritten as follows

$$
P_{\rm Ps} = k_{ep} \int_0^\infty {n_e n_p \over \tilde
             V_{ep}} dt = k_{ep} \int_0^\infty
\frac{n_e n_p}{V^0_{ep} (1+t/\tau_{ep})^{3/2}}
\exp \left[ -\mu_{ep}^2 \cdot
     \frac{(\alpha +t/\tau_{ep})^2}{1+t/\tau_{ep}} \right]
    \cdot dt,
\eqno(18)
$$
where
$$
V^0_{ep} = [\pi (a^2_{bl}+a^2_p)]^{3/2},
\qquad
\tau_{ep} = {a^2_{bl}+a^2_p \over 4(D_{amb} + D_p)},
$$
$$
\mu_{ep}^2 = {v_p^2 \tau_{ep}^2 \over a^2_{bl}+a^2_p }
           = \left( \frac{e D}{4\varepsilon T} \right)^2
           (a^2_{bl}+a^2_p),
\quad \alpha = \frac{2}{3}\cdot \frac{a^2_p-a^2_{bl}}{a^2_p+a^2_{bl}}.
$$

If we neglect positron annihilation and localization, i.e. put $\tau_p^{-1} = 0$, one obtains

$$
P_{\rm Ps} = 1 - n_p(\infty).
\eqno(19)
$$
This equation is a conservation law for positrons: $1- n_p(\infty)$ is the fraction of Ps
formation and the remaining $n_p(\infty)$ is the fraction of positrons annihilating in a "free"
state at a time scale on the order of $\tau_2 \equiv \lambda_f^{-1}$.

Application of the prescribed diffusion method to solve Eqs.(10) is based on an assumption that interaction between thermalized
positron and its blob is negligible. In other words, we adopted that the blob is "transparent" for $\rm e^+_{qf}$ (we call this
approach the "white blob" model).  This assumption seems reasonable because of two opposite effects which approximately cancel
each other (Sec.IV.G):  1) outdiffusion of intrablob electrons makes them to reside in an outer region of the blob, which
results in appearance of an excess positive charge in its central region, repelling e$^+$ from the blob; 2) the presence of
e$^+$ within the blob may lead to rearrangement of intrablob electrons, which may decrease the total energy of the system
because of the Debye screening.

Thus, in this case fast diffusion motion of the quasifree positron on a timescale of its lifetime (as we shall see in Sec.IV.E
$\tau_{ep} \ll \tau_2$) efficiently smears the positron distribution and approaches it to a Gaussian shape. This justifies
application of Eqs.(14)-(15).

\subsection{The simplest case of Ps formation. Relation to the Onsager approach}

If the Ps formation process is assumed to be very fast compared to IER, possible localization
processes and diffusion expansion of the blob ($k_{ep} \gg k_{ie} \to 0 $, $\tau_e \to \infty$,
$\tau_p \to \infty $ and $D_p \gg D_{amb}$), solutions of Eqs.(10) become

$$
n_e(t< \tau_e) = n_0,  \qquad
n_p(t< \tau_p) = \exp \left[ - W_{ep} \int_0^{t/\tau_{ep}}
            {d\vartheta \over (1+ \vartheta)^{3/2} }
\exp \left( - \mu_{ep}^2 {(\alpha +\vartheta)^2 \over
              1+\vartheta} \right) \right],
\eqno(20)
$$
where
$$
W_{ep} = \frac{n_0 k_{ep} \tau_{ep}}{V_{ep}^0} =
         \frac{n_0 k_{ep}}{4\pi
(D_{amb}+D_p)\sqrt{\pi(a^2_{bl}+a^2_p)}} \approx  \frac{n_0
k_{ep}}{4\pi          D_p \sqrt{\pi(a^2_{bl}+a^2_p)}}.
\eqno(21)
$$
is a dimensionless parameter that integrally accounts for such factors as diffusion of the positron,
e$^+$-e$^-$ electrostatic attraction, the efficiency of the absorption of released energy, free volume and
its distribution, which determine the Ps formation probability in a zero field. Substituting Eq.(20) into
Eq.(19), we obtain

$$
P_{\rm Ps} = 1 - \exp \left[ - W_{ep} \int_0^\infty
             {d\vartheta \over (1+ \vartheta)^{3/2} }
\exp \left( - \mu_{ep}^2 {(\alpha +\vartheta)^2 \over
              1+\vartheta} \right) \right],
\eqno(22)
$$
$$
\mu_{ep} = {\sqrt{a^2_{bl}+a^2_p} \over 4} \cdot
           {e D\over \varepsilon T}
         = 10^{-4} \sqrt{a^2_{bl}+a^2_p} \hbox{ (\AA)} \cdot
           {D \over \varepsilon} \left( {\hbox{kV} \over
            \hbox{cm}} \right).
\eqno(23)
$$
In the present experiments at room temperature, the highest experimentally reachable field was
$D=25$ kV/cm. If $\sqrt{a^2_{bl}+a^2_p} \sim 100$ \AA, $\mu_{ep}$ is about 0.1. It is worth
mentioning that the drift $v_p \tau_{ep}= \mu_{ep} \sqrt{a_{bl}^2 +a_p^2}$ of the positron
distribution during Ps formation time is small in comparison with the e$^+_{\rm qf}$ diffusion
distance $\sqrt{a_{bl}^2+a_p^2}$ during the same time.

At a small $\mu_{ep}$ an asymptotic expression of Eq.(22) is:

$$
P_{\rm Ps}(\mu_{ep}\ll 1) = 1-\exp \left[ -2W_{ep}
  \left( 1-\sqrt{\pi}\mu_{ep} + {\mu_{ep}^2\over 3}(8-5\alpha-
                            \alpha^2)
   - \ldots \right) \right].
\eqno(24)
$$
It is seen from Eq.(24) that $\alpha$ only slightly affects the $P_{\rm Ps}$
through the highest orders of $\mu_{ep}$. In a zero field ($\mu_{ep}=0$):

$$
P_{\rm Ps}(0) = 1-\exp (-2W_{ep}).
\eqno(25)
$$

Note that this equation reproduces the form of the well-known Onsager
result for geminate recombination, which in low fields is written as:
\cite{Hum74}

$$
P_{\rm gr} = 1-\left(1+ {eD r_c \over 2T} \right) \cdot
         \exp \left(-{r_c \over r_0}\right) ,
    \qquad
        r_c = {e^2 \over \varepsilon T},
\eqno(26)
$$
where $r_0$ is the initial separation of the geminate e$^+$-e$^-$ pair.  If
$2\sqrt{\pi}W_{ep}\mu_{ep} \ll 1$, we can expand the exponent in Eq.(24) and keep only the leading
field-dependent term:

$$
P_{\rm Ps} = 1- \left(1+ {e D \tilde r_c \over 2\varepsilon T}\right)
            \cdot
        \exp\left(-{\tilde r_c\over \tilde r_0}\right) ,
\qquad
    \tilde r_c = \frac{n_0 k_{ep}}{4\pi D_p},
\qquad
    \tilde r_0 = \frac{\sqrt{\pi (a^2_{bl}+a^2_p)}}{2}.
\eqno(27)
$$
The meaning of  radius $\tilde r_c$ in Eq.(27) differs from the Onsager
radius  used in Eq.(26). Eq.(26) corresponds to the $\delta$-function
distribution of $e^+$-$e^-$ initial separation, $\delta (r-r_0)$, while
Eq.(27) is obtained assuming Gaussian distributions for the positron and
electron.

Note that Eq.(26) contains only one fitting parameter, $r_0$, while Eq.(27)
has two. We thus have more freedom to fit experimental data with Eq.(27).
With Eq.(26), we can fit only relative Ps yield $P_{\rm Ps}(D)/P_{\rm
Ps}(0)$. \cite{Ito79,Ito88,Wan98} This is an important difference between
the Onsager model (spur model) and the blob model. It is also worth
mentioning that the dielectric permittivity enters Eq.(26) and Eq.(27) in a
different way.

For a large $\mu_{ep}$, the asymptotical expansion of Eq.(22) becomes

$$
P_{\rm Ps}(\mu_{ep}\gg 1) = 1 - \exp \left[
           - {\sqrt{\pi} W_{ep} \over 2\mu_{ep} }
    \left(1-{1\over 2\sqrt{\pi}\mu_{ep}} + \ldots  \right)\right].
\eqno(28)
$$
It means that $P_{\rm Ps}$ should go to zero at high fields.  However, this
regime is definitely beyond the applicability of the present theory (for
example, inequality $\mu_{ep}> 5$ implies that $D$ should be larger than
$10^3$ kV/cm). In this case another Ps formation process comes into play
(Sect. III B).

\subsection{e$^+$ annihilation and ion-electron recombination}

A more realistic consideration than that in the previous section must take into account at least e$^+$ annihilation and the
possibility of IER (reaction $8c$). Both of these processes may compete with Ps formation. IER equally decreases the number of
electrons and ions. If we neglect as before the diffusion expansion of the blob,

$$
n_e(t) = n_i(t) = \frac{n_0}{1+ n_0 k_{ie} (2\pi a_{bl}^2)^{-3/2} t}.
\eqno(29)
$$
In this case integration of Eq.$(16c)$ gives

$$
n_p(t) = \exp \left[ - {t\over \tau_p} - W_{ep}
         \int_0^{t/\tau_{ep}} {d\vartheta \over (1+
         \vartheta)^{3/2} (1+W_{ie}\vartheta)}
      \exp \left( - \mu_{ep}^2 {(\alpha+\vartheta)^2 \over
                      1+\vartheta} \right)
              \right],
\eqno(30)
$$
where
$$
W_{ie} = \frac{n_0 k_{ie} \tau_{ep} }{ V_{ie}^0 }
\approx  \frac{n_0 k_{ie}(a^2_{bl}+a^2_p)}{4 D_p (2\pi
a^2_{bl})^{3/2}}.
\eqno(31)
$$
Unfortunately, because of the presence of the decay term $n_p/\tau_p$ in
Eq.$(16c)$ we cannot use the relationship $P_{\rm Ps}= 1-n_p(\infty)$ as
before and must use Eq.(18) for calculation of $P_{\rm Ps}$. Its integration
by parts gives
$$
P_{\rm Ps} =
   1 - W_p \int_0^\infty d\vartheta
   \exp \left[ - W_p\vartheta - W_{ep} \int_0^\vartheta
   \frac{dz \exp \left( -\mu_{ep}^2 \frac{(\alpha+z)^2}{1+z}
                    \right) } {(1+z)^{3/2} (1+W_{ie}z)}
        \right]=
$$
$$
=   1 - \int_0^\infty d\vartheta
   \exp \left[ - \vartheta - W_{ep} \int_0^{\vartheta/W_p}
   \frac{dz \exp \left( -\mu_{ep}^2 \frac{(\alpha+z)^2}{1+z}
                    \right) } {(1+z)^{3/2} (1+W_{ie}z)}
        \right],
\eqno(32)
$$
$$
W_p = \frac{\tau_{ep}}{\tau_p} = \frac{a^2_{bl}+a^2_p}{4 D_p}
       \left( \frac{1}{\tau_2} + \frac{1}{\tau_p^{loc}} \right).
\eqno(33)
$$
A distinctive feature of this relationship is a behavior at low fields. It is not a linear
decrease with $D$ as it was in Eq.(22), Eq.(24) and Eq.(27), but a quadratic one.  At non-zero
$W_p$ or $W_{ie}$, straightforward expansion of the exponent $\exp\left( - \mu_{ep}^2
{(\alpha+z)^2 \over 1+z} \right) \approx 1 - \mu_{ep}^2 {(\alpha+z)^2 \over 1+z} $ in small
fields in Eq.(32) leads to
$$
P_{\rm Ps}(D=0) - P_{\rm Ps}(D\to 0) \propto D^2.
\eqno(34)
$$
This important peculiarity is a result of competition between the Ps formation process, IER and e$^+$ annihilation (or
localization.

\section{Comparison with experimental data and discussions}

\subsection{$P_{\rm Ps}(D)$: Simplest consideration}

Here we consider experimental data of Kobayashi and co-workers obtained by means of the
positron annihilation lifetime spectroscopy in the following nonpolar liquids: benzene,
hexafluorobenzene, hexane, cyclohexane, and isooctane. \cite{Ste00,Wan98} Air dissolved in the
liquids was removed by the standard freeze-thaw method.  High voltages were applied to liquids
through a pair of electrodes to yield external fields of up to 25 kV/cm. In a given electric
field, one lifetime spectrum was collected for 2.5 hours at room temperature (295 $\pm$2 K),
resulting in a total of $\sim 7\cdot 10^5$ counts for each spectrum.  Measured lifetime
spectra were decomposed into three components.  The first and second components,
$\tau_1=$190-310 ps and $\tau_2$ (Table I), were assigned to the annihilation of p-Ps and
free positrons.  The longest-lived component, $\tau_3$, was due to the ortho-Ps (o-Ps)
annihilation in bubbles in liquids (Table I). The o-Ps lifetime was unchanged by an increase
in the external electric field.

To compare our theoretical results with experimental data, we relate $P_{\rm Ps}$ to the
intensity of o-Ps component $I_3$.  We assume that $I_3=3P_{\rm Ps}/4\cdot 100\%$. The
multiplier 3/4 in this relation may vary slightly  from one liquid to another because of
possible Ps interactions with highly reactive radiolytic products (localized electrons,
radicals, and radical-cations). \cite{Mog95,Bya96}

Experimental data are shown in Fig.1. As a first step, we tried to fit experimental data
to Eq.(22), adjusting two parameters: $\sqrt{a_{bl}^2+a_p^2}$ and $W_{ep}$
(Fig.\ref{I3_D} and Table II). Parameter $\alpha$ ~($0<\alpha<2/3$) entering Eq.(22) was
not free. It was recalculated from $\mu_{ep}$ assuming $a_{bl}=40$ \AA. It is seen from
Eq.(24) that $\alpha$ has small influence on $P_{\rm Ps}$.  In all liquids investigated
but C$_6$F$_6$, $\sqrt{a_{bl}^2+a_p^2} $ is 100-200 \AA. Obtained e$^-$-e$^+$
recombination rate constants (Table II) are of order of IER rate constants measured in
pulse radiolysis experiments in radiation chemistry: $4.7\cdot 10^{13}$-$7.2\cdot
10^{13}$ M$^{-1}$s$^{- 1}$ in n-hexane, $1.9\cdot10^{14}$ M$^{-1}$s$^{-1}$ in cyclohexane
and $\sim 2\cdot 10^{15}$ M$^{-1}$s$^{-1}$ in isooctane. \cite{Hat94}  As mentioned,
$k_{ep}$ includes many different factors, e.g., e$^+$-e$^-$ electrostatic attraction,
availability of the appropriate free volume, and the absorption of the released energy,
etc.

\subsection{Effect of the positron annihilation and ion-electron recombination }

\begin{figure}[t]
\noindent
    \epsfig{file=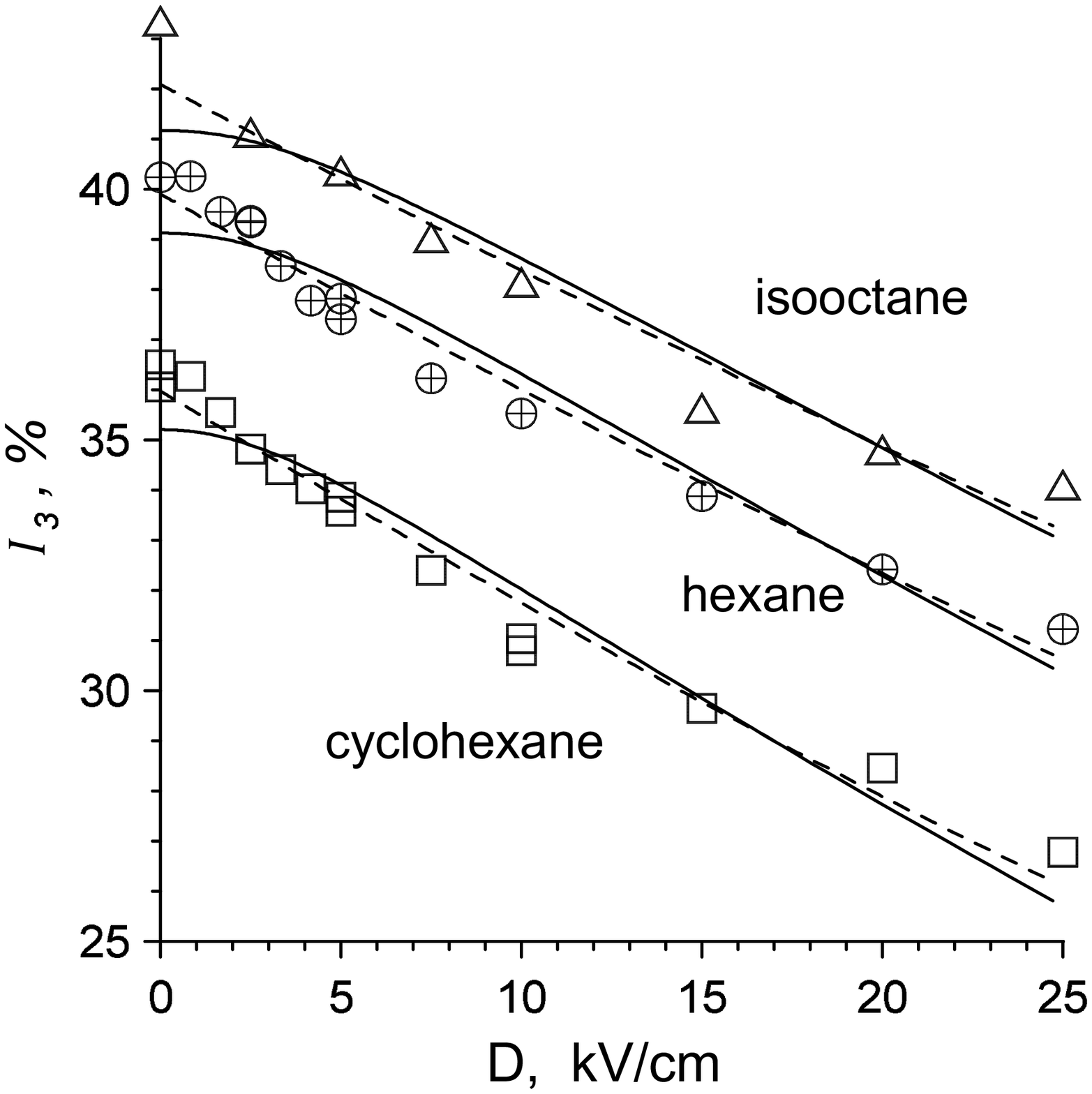, width=75mm}
        \hspace{3mm}
    \epsfig{file=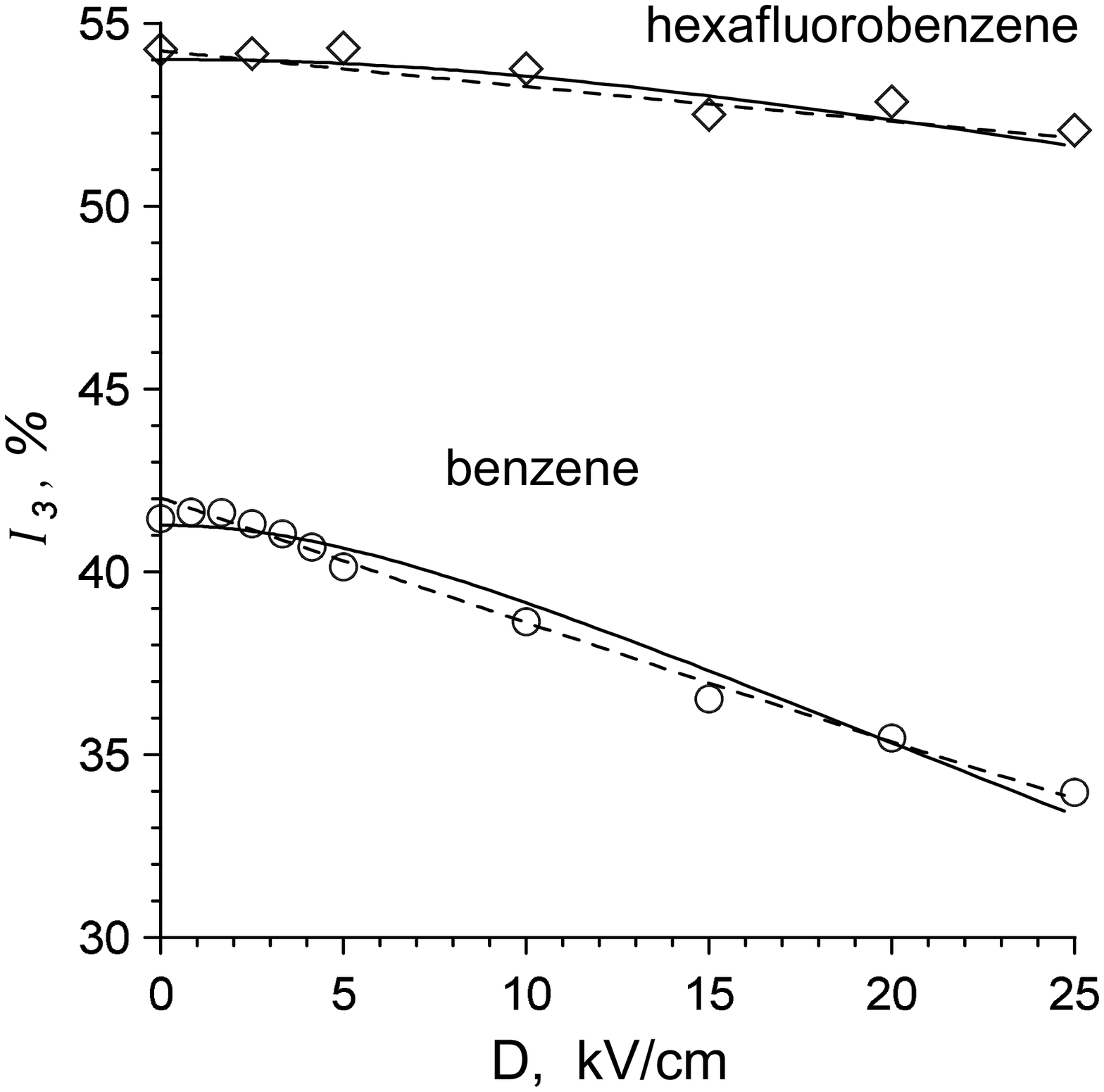, width=75mm}
\vspace{1cm} \caption{}\label{I3_D}
\end{figure}

\begin{figure}
    \begin{center}
    \epsfig{file=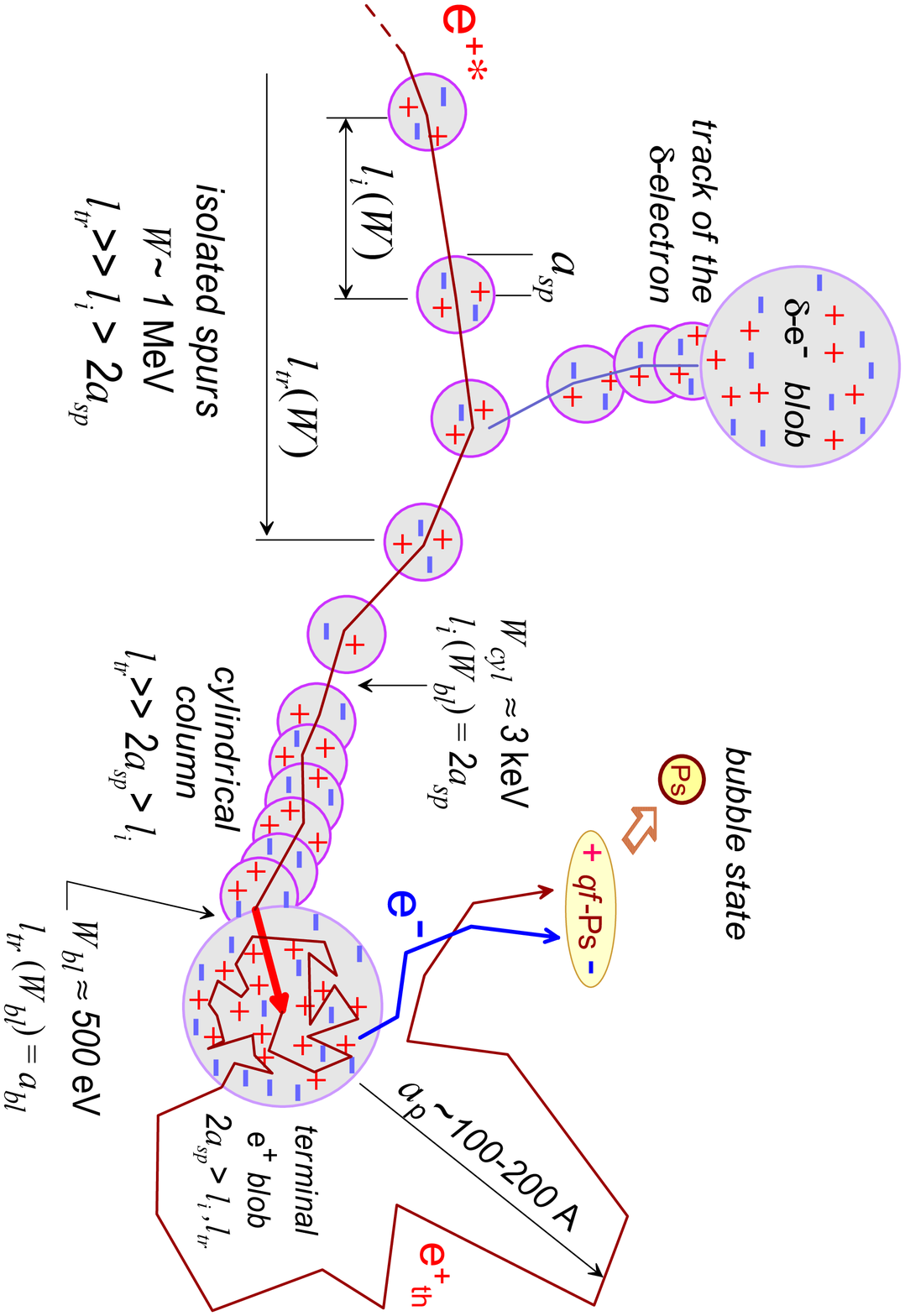, width=15cm, angle=0}
    \end{center}
    \vspace{2cm} \caption{} \label{e_track}
\end{figure}

It is seen that the electric field effect is stronger in saturated hydrocarbons (cyclohexane, hexane, isooctane) than in
aromatic compounds. Experimental data in Fig.\ref{I3_D} in benzene and especially in hexafluorobenzene indicate that the slope
of $P_{\rm Ps}(D)$ tends to zero as the field approaches zero. This observation is reasonable if Ps formation competes with
e$^+$ annihilation, e$^+$ localization, or IER. Treatment of the data based on Eq.(32) (solid lines in Fig.\ref{I3_D}) with
$W_{ie}$=0 and $1/\tau_p^{loc}$=0 leads to a zero slope of $P_{\rm Ps}({\rm D})$ at D=0 and larger values of parameter
$\sqrt{a_{bl}^2+a_p^2}$ (Table II). The introduction of non-zero values of $W_{ie}$ worsens the fit, but if we allow that
$W_{ie}/W_{ep}$ is 0.005, final curves coincide practically with solid lines, that is why we did not plot them.  The values of
fitting parameters are listed in Table II. Based on these results,  the maximal limiting value of the $W_{ie}/W_{ep}$ ratio was
found to be 0.005. Using the numerical values of $a_{bl} \approx 40$ \AA\ and $n_0 \approx 30$, the ratio of rate constants
$k_{ie}/k_{ep}$ turns out to be very small, about $10^{-3}$ (Table II), which implies that IER cannot compete significantly
with Ps formation; note, however, that the values of $k_{ie}$ estimated here do not correspond to typical experimental
conditions in radiation chemistry.  This indicates that, in the terminal positron blob, Ps formation proceeds within a shorter
time than IER. The following three points make this clear:

1. Both Ps formation and IER reactions depend strongly on the same process -- energy
absorption and the transfer of released energy. As discussed in Sect. II D.4, because of the
high density of ion-electron pairs and quasi-neutrality condition in the blob, thermalization
and IER are lengthened.  Ps formation proceeds more easily, however, because both particles
lose energy and the total released energy is less than that in IER.

2. $k_{ie}$ is proportional to the overlapping of wave functions of a delocalized state
(quasifree electron) and a localized state (positive ion), while $k_{ep}$ is proportional to a
much larger overlapping of the two delocalized wave functions of the quasifree positron and
electron.

3. As mentioned above the mobility of $\rm e^+_{qf}$ is higher than that of $\rm e^-_{qf}$
(Table I), leading to a larger ratio, $D_p/D_e$, proportional to $k_{ep}/k_{ie}$.

\subsection{e$^+$ localization}

To avoid the strong disagreement with experimental data, we must assume that e$^+$ localization
does not proceed in neat liquids.  $\tau_p^{loc}$ at least should be longer than $\tau_2$ (Table
II), which agrees well with our previous conclusion (Sect. II E.2) that e$^+$ localization is
unfavorable in neat nonpolar liquids other than C$_6$F$_6$.

\subsection{C$_6$F$_6$}

The C$_6$F$_6$ molecule may capture an epithermal electron in a shallow energy level. \cite{Gan76} "Shallow" implies that the
positron may pick up an electron from C$_6$F$_6^-$ and form a Ps atom.  Trapped electrons escape IER but survive for Ps
formation by remaining on a molecule of the liquid. Localization of intrablob electrons decreases the $\psi$-function
overlapping with $\rm e^+_{qf}$, which results in a smaller rate constant $k_{ep}$ (Table II). Product
$D_p\sqrt{a_{bl}^2+a_p^2}$ also turns out to be small, however, $W_{ep}$ and therefore the Ps formation probability at zero
field ($P_{\rm Ps}(0) \approx 1-e^{- 2W_{ep}}$, Eq.(25)) is the largest among the liquids investigated. The small value of the
positron thermalization distance in C$_6$F$_6$ qualitatively agrees with the data of Gee and Freeman for the thermalization
distance of the excess electron. \cite{Gee90}

Why is $D_p \sqrt{a_{bl}^2+a_p^2}$ small in hexafluorobenzene? The higher electron density $n_e$ and the larger number $Z$ of
electrons in C$_6$F$_6$ (Table I) enhance positron scattering and decrease $D_p \propto 1/Z n_e$. Thus e$^+$ mobility $b_p $
drops below the experimentally detectable limit. \cite{Wan00}

\subsection{Ps formation time}

Using the numbers for $\sqrt{a_{bl}^2+a_p^2}$ and adopting approximate positron mobility $b_p$
equal to 10 cm$^2$V$^{-1}$s$^{-1}$ (Table I), which corresponds to $D_p=0.25$ cm$^2$/s, we
obtain  Ps formation time $\tau_{ep} \approx (a_{bl}^2+a_p^2) /4D_p$ on the order of some
picoseconds for the liquids except C$_6$F$_6$ in accordance with other estimations.
\cite{Mil86,Wan98}

\subsection{ Absence of the field dependence of $\tau_3$ }

Experimental data clearly show that lifetime $\tau_3$ of the o-Ps atom residing in a bubble, does not depend on applied
electric field within ranges of statistical deviations in all investigated liquids.\cite{Wan00} It is well-known that the o-Ps
annihilation rate is proportional to the overlapping of the positron wave function with that of the electrons of the molecules
("pick-off" process).

The Ps atom is electrically neutral, but a highly polarizable system.  Nevertheless an external
field $\rm D \approx 25$ kV/cm stretches e$^+$ and e$^-$ in Ps only by a distance about $\sim
10^{-5}$\AA. It slightly increases the overlapping in one side of the Ps bubble and decrease it by
the same amount on the opposite side. Total overlapping remains constant.  This is the reason for
the absence of the field dependence of $\tau_3$ values.

\subsection{Electrostatic and polarization effects}

More careful inspection of our experimental data at high fields and other related experimental
results \cite{Wan98,Ito88,Mog75} suggests that Ps formation probability tends to reach a
plateau.  It in fact could be attributed to a nontrivial role of polarization interaction between
the thermalized quasifree positron and the blob, which is not yet well understood.

Being highly mobile, intrablob electrons tend to reside in an outer region of the blob (Appendix
B). It results in the appearance of a small excess positive charge in its central region, which
repels e$^+$ outward. This potential, Eq.(B7), is about $3T$ in the center of the blob and
decreases to zero at $r\gtrsim a_{bl}$.

Polarizability $\alpha_{bl}$ of the blob as a whole is about $\varepsilon a_{bl}^3$, so at
$r>a_{bl}$, the e$^+$-blob polarization attraction is about $-\frac{\alpha_{bl}}{2}
\frac{e^2}{\varepsilon^2 r^4}$, where $r$ is e$^+$ separation from the center of the blob. Its
maximum possible value (at $r\sim a_{bl}$) is about $-\frac{e^2}{\varepsilon a_{bl}} \sim -0.1$
eV, which is larger than $T$, so this shallow negative potential in the outer region of the blob
may trap the thermalized positron.

Because of the presence of the large number of ion-electron pairs inside the blob, e$^+$ potential energy decreases by a value
$\sim \frac{-e^2}{\varepsilon (r_D + a_{bl}/n_0^{1/3})}$ due to correlations in positions of the charged intrablob species,
which is, in essence, Debye-Huckel screening of the positron charge. Here the Debye radius $r_D \approx (4\pi r_c
c_{iep})^{-1/2} \approx 4$ \AA, where $c_{ier} \approx n_0 \left/ \frac{4}{3}\pi a_{bl}^3 \right.$ is the concentration of
ion-electron pairs in the blob, and $a_{bl}/n_0^{1/3}$ is the average distance between intrablob species. At the distances less
than the average distance between the particles $r\lesssim a_{bl}/n_0^{1/3}$ the screening potential is pure Coulombic. At
larger $r$ it takes the Debye form.  Such screening of the positron by blob charges also makes residence of the thermalized
positron inside the blob favorable.

These effects are rather subtle. Quantitatively, they are somewhat higher than $T$ ($\sim 0.1$
eV), but have different signs in $r$-space. Nevertheless it is possible that $\rm e^+_{qf}$ may be
trapped inside or near the blob.  This circumstance is important for interpretation of the
electric field effect on Ps formation and may be related to a tendency of $P_{\rm Ps}$ reaching
"plateau" at high fields.

The approach developed here properly takes into account the presence of the blob as an inhomogeneously distributed large number
of ion-electron pairs and assumes that all of the above polarization effects more or less compensate each other. More accurate
consideration of these electrostatic and polarization effects of the e$^+$-blob interaction is the subject of our following
study (so-called "black blob" model). \cite{Ste01MSF}

\section{Conclusion }

The difference between radiation chemistry and Ps chemistry is related to the difference in the objects they study.  Being a
probe of Ps chemistry, the positron delivers information about processes near and inside its terminal blob, while radiation
chemistry primarily investigates the processes in isolated spurs. The main difference between the spur and blob comes from the
factor of 10 difference in the initial number of ion-electron pairs they contain.

In radiation chemistry, Onsager's theory of the geminate recombination is adequate for
interpreting free-ion yields and their field dependence. On the contrary, the blob model is most
appropriate for considering processes, in particular, Ps formation in the blob.  At a limiting
case, the theory we have developed gives a relation which only formally resembles the well-known
Onsager equation of geminate recombination.  We generalized the prescribed diffusion method for
consideration of the processes in the presence of an external electric field.

Thermalization of knocked-out electrons differs from the slowing of the positron.  Being
affected by an electric field of the parent ion, the electron becomes thermalized at a shorter
distance than the positron, which easily escapes from its electrically neutral center of the
blob. Thus $a_p$ becomes larger than the initial ion-electron separation, $r_0$, in a spur.
This is also related to the higher mobility of the thermalized positron compared to the electron
mobility.

An investigation of Ps formation in the presence of an external electric field enables us to
better understand the behavior of intrablob electrons and the positron -- peculiarities of
their thermalization, quasi-neutrality condition, positron out-diffusion from the blob,
formation of the quasifree positronium and its transformation to the bubble state.

One of the interesting findings in our study is that the competition of the Ps formation
with the other processes (like annihilation of positrons in the free state and IER) leads
to a zero derivative of $P_{\rm Ps}(D)$ at $D=0$. Another surprising but not very
unexpected observation is that IER cannot compete significantly with the Ps formation. We
found that the ratio of the Ps formation rate constant to the IER rate constant
$k_{ep}/k_{ie}$ is about $10^3$-$10^4$. Different conditions of absorption of released
energy in IER and Ps formation in the e$^+$ blob may be responsible for this result.

In summary, we stress that experiments on Ps formation in the presence of an external electric
field are extremely informative both for Ps chemistry and for radiation chemistry, and should be
extended to higher fields and other substances.

\vspace{5mm}
{\bf Acknowledgments}
\vspace{5mm}

The authors are indebted to Dr. Y.Kobayashi, Dr. Cai-Lin Wang and Dr. K.Hirata for doing additional measurements at low
electric fields, discussions and hospitality during our stay in Tsukuba. We thank Professors  David M. Schrader and Tomasz
Goworek for useful comments.

This work was undertaken as a part of the Nuclear Cross-Over Research Program with the financial
support of the the Science and Technology Agency (STA) of Japan which enabled Sergey Stepanov to
work at the National Institute of Materials and Chemical Research. Acknowledgment is made to the
Russian Foundation of Basic Research (Grants 00-03-32918 and 00-15-96656).

\appendix
\section{Energy deposition and track structure of fast positron}

The average initial energy of fast positrons emitted from radioactive nuclei $^{22}$Na or $^{64}$Cu, frequently used as a
source of positrons, is several hundreds of keV. Moving in a liquid, the positron loses about  half of its kinetic energy in
rare head-on collisions, knocking out $\delta$-electrons.  Tracks of these electrons form "branches" around the positron trail.
The other half of energy is spent in numerous glancing collisions with molecules. Average energy loss in such a collision is
30-50 eV (at maximum 100 eV).  A secondary electron knocked out in a glancing collision produces, by turns, a few ion-electron
pairs inside a spherical microvolume, called a spur in radiation chemistry. Its radius, $a_{sp}$, is determined by
thermalization of the knocked out electrons in the presence of the Coulombic attraction of parent ions.  Strictly speaking,
$a^2_{sp}$ is the dispersion of the Gaussian distribution function, which corresponds to the end of thermalization process of
the knocked-out e$^-$.  Based on \cite{Ste93,Ste95}, the most probable value of $a_{sp}$ in water is estimated to be $\approx
30$ \AA. \cite{a_sp}

While positron energy $W>W_{cyl}$, mean distance $l_i$ between adjacent events of ionization produced by the positron is
greater than spur size $2a_{sp}$ (Fig.\ref{e_track}). \cite{Bya96} This means that spurs are separated from each other.  The
motion of a high energy positron is a quasi-straight line because $l_i$ is less than the positron transport path $l_{tr}$,
which is the mean distance passed by the positron before it changes the initial direction of its motion by 90$^\circ$ (detailed
in \cite{Jab98}). When $l_i<2a_{sp}<l_{tr}$ or $W_{bl}<W<W_{cyl}$ spurs overlap, forming something like a cylindrical
ionization column.  At the end of the track, when e$^+$ energy becomes less than blob formation energy $W_{bl}$ ($\sim 500$
eV), $2a_{sp}$ becomes the largest parameter: $2a_{sp}>l_i, l_{tr}$. This means that the positron starts to create a blob
containing a few tens of ion-electron pairs because the average energy required for the formation of one ion-electron pair is
16-22 eV.\cite{Bya85}  Diffusion motion of the positron in the blob becomes more pronounced with decreasing energy.  The
positron frequently changes its momentum due to elastic scattering and the ionization of surrounding molecules. Roughly
speaking all intrablob ionizations are confined within the sphere of the radius $a_{bl}$ (detailed definition see below). The
positron finally becomes subionizing and therefore its energy loss rate drops down by almost 2 orders of magnitude.
\cite{Ste95}

\begin{figure}
    \begin{center}
    \epsfig{file=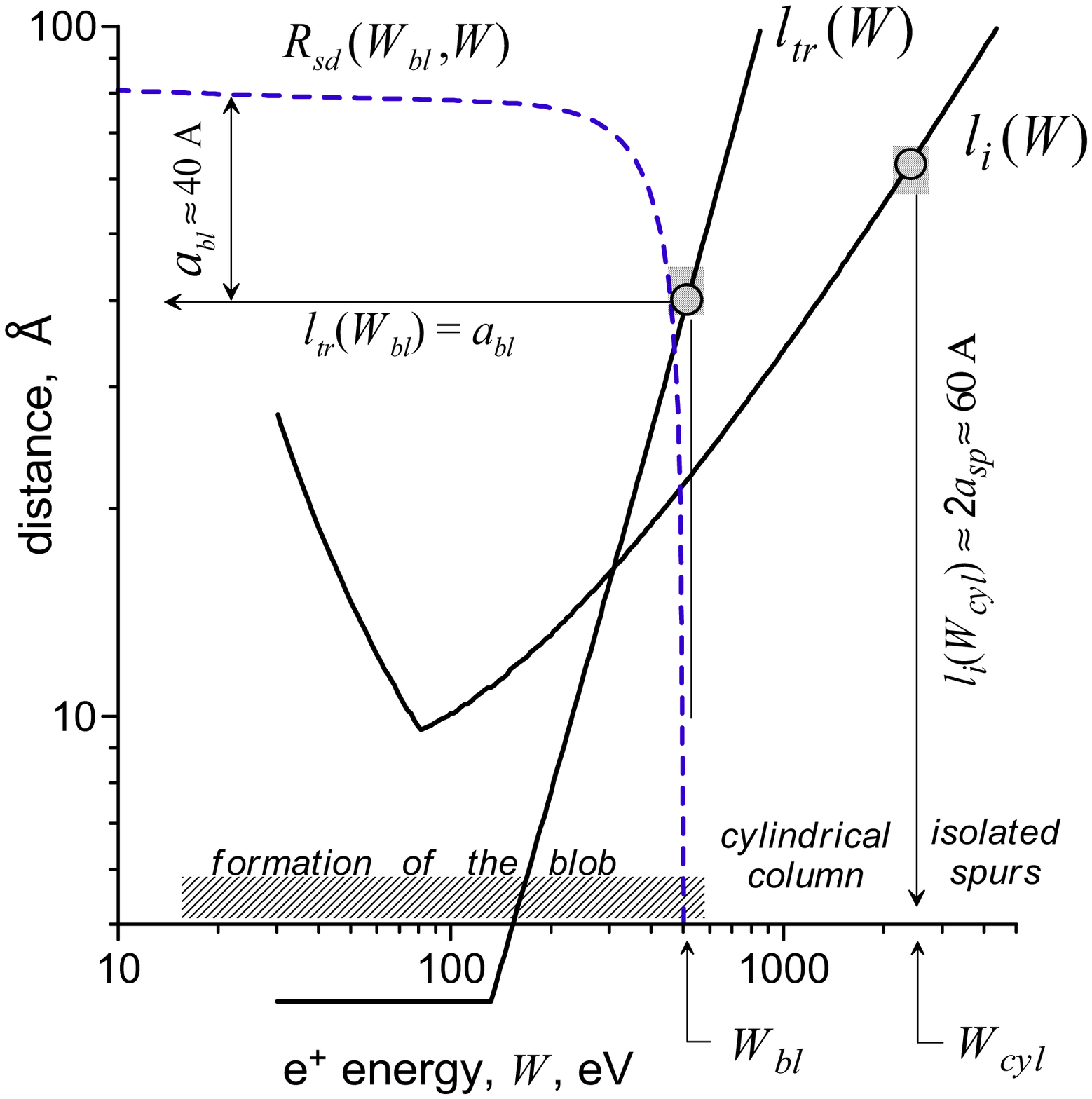, width=16cm}
    \end{center}
\vspace{1cm} \caption{}\label{a_bl_W_bl}
\end{figure}

To gain some insight into typical values of the parameters involved we shall start with estimations of $l_i(W)$ and
$l_{tr}(W)$. Calculation of $l_i(W)=W_{iep}/{\rm LET}(W)$ is based on the data on LET of e$^\pm$ (see, for example, Fig.3.19 in
\cite{Par87} or Fig.5 in \cite{Bya96}). Usually the right side of the Bragg peak is well described by the Bethe formula for
ionization slowing down, but its left (low energy) side strongly depends on corrections to the Bethe equation which consist in
truncation of the dipole oscillator distribution at the maximum transferable energy.\cite{Moz72} From the relationship
$l_i(W_{cyl})\approx 2a_{sp}$ we obtain that $W_{cyl}\approx 3$ keV, Fig.\ref{a_bl_W_bl}.

Estimation of the transport path can be done in the frameworks of the Born approximation (wavelength of e$^+$ with the energy
$\gtrsim 100$ eV is small in comparison with the size of molecules). Below 1 keV $l_{tr}(W)$ is mainly determined by
small-angle electron-positron elastic scattering (the Rutherford part of the cross-section, related to e$^+$ scattering on
nuclei, becomes important above 1 keV): \cite{Bya96}
$$
l_{tr} (W) = {1 \over n \sigma_{tr}(W)},
    \qquad
\sigma_{tr}(W) =
    \int_0^\pi |f_B(\theta)|^2 (1-\cos \theta)
                                2\pi \sin \theta d \theta. \eqno(A1)
$$
The Born amplitude $f_B$ is calculated simulating a molecule of the liquid by an equi-electronic
hydrogen-like atom \cite{Bya85,Bya96}. Energy dependence of $l_{tr}$ is shown in
Fig.\ref{a_bl_W_bl}.

At low energies positron scattering becomes more and more efficient and we must regard e$^+$ motion as
diffusion-like. $l_{tr}$ is then considered as the energy-dependent mean free path between successive
"collisions", which completely randomize the direction of the velocity of the particle. If the probability to
pass distance $r$ without such a collision is $\exp(-r/l_{tr})$, the average squared distance
$\overline{r^2}$ is $2 l_{tr}^2$. After $n$ collisions, the mean square displacement is $n \overline{r^2}$.
Calculation of the same quantity assuming that particle propagation is governed by the usual diffusion
equation gives $n \overline{r^2} = 6D_p t$. Thus, we obtain the diffusion coefficient of the positron as
$D_p(W) = l_{tr}^2/3\tau = l_{tr} v_p/3$, where $v_p$ is e$^+$ velocity and $\tau = t/n = \bar r/v_p =
l_{tr}/v_p$ is the average time between subsequent collisions.  Integrating the relationship
$$
d(r^2) = 6D_p dt = 2 l_{tr} v_p dt = 2 l_{tr} dx =
    2 l_{tr} \frac{dW}{|dW/dx|_{ion}}
\eqno(A2)
$$
from the energy $W_i$ down to $W_f$, we obtain the diffusion displacement $R_{sd}$ of e$^+$ during its ionization slowing down
within this energy interval $W_f<W<W_i$:
$$
R_{sd}(W_i,W_f) = \left( 2 \int_{W_f}^{W_i} l_{tr}(W) \frac{dW}{|dW/dx|_{ion}}
                  \right) ^{1/2}.
\eqno(A3)
$$

\begin{figure}
    \begin{center}
    \epsfig{file=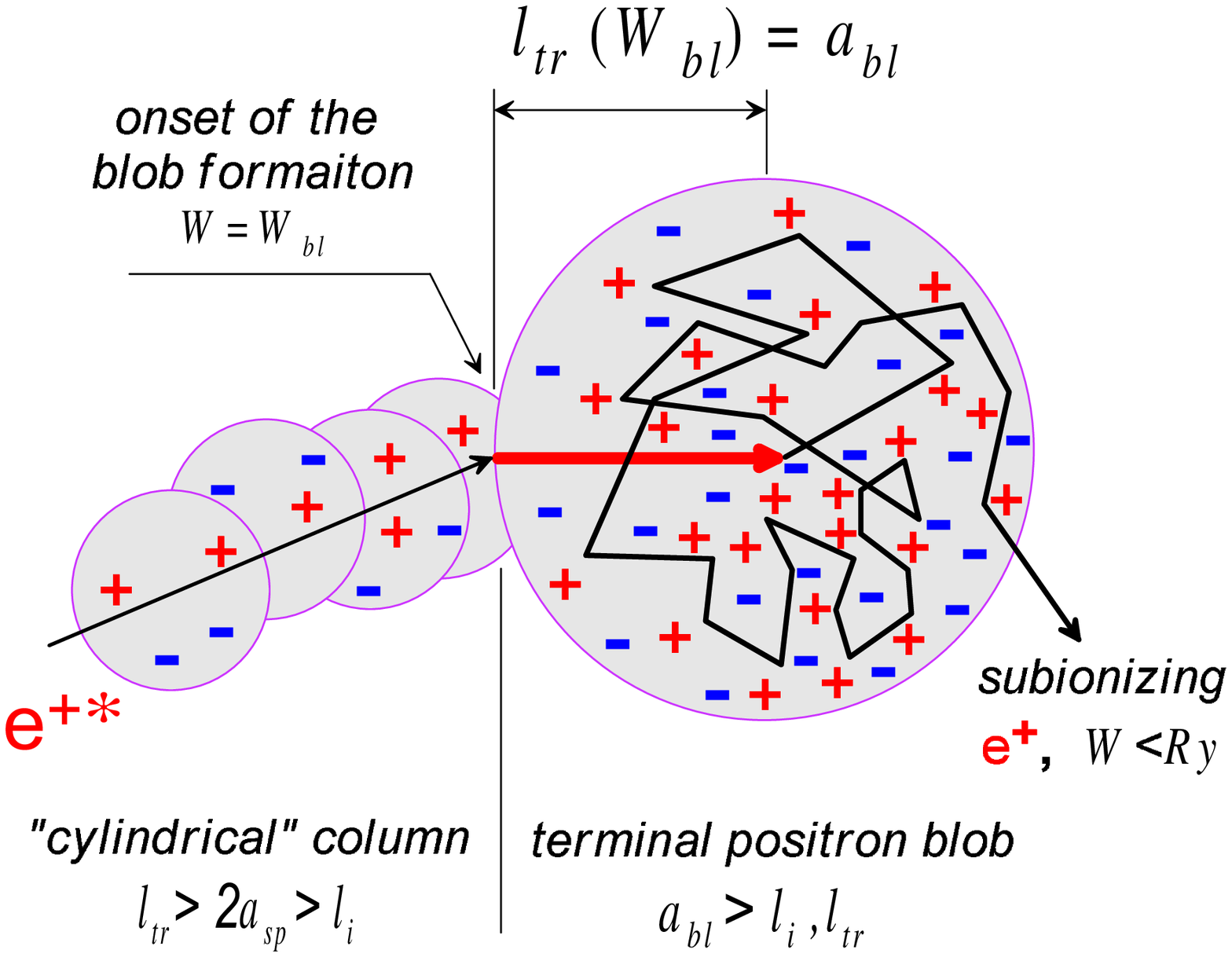, width=15cm}
    \end{center}
\vspace{1cm} \caption{}\label{bl_sch}
\end{figure}

Now we are ready to define the blob formation energy $W_{bl}$ and the "radius"\ $a_{bl}$ of the blob. These
quantities are determined from the following equations:
$$
l_{tr}(W_{bl}) = a_{bl},
    \qquad
 a_{bl} = R_{sd}(W_{bl},Ry) - a_{bl}.
\eqno(A4)
$$
Their physical meaning is clear from Fig.\ref{bl_sch}. Terminal blob is a spherical microvolume which
confines the end part of the positron trajectory, where ionization slowing down is the most efficient
(thermalization stage of subionizing positron is not included here). Mathematical formulation of this
statement is twofold. Just after the first blob formation "step" (the thick arrow in Fig.\ref{bl_sch}),
$l_{tr}(W_{bl})$) e$^+$ reaches the center of the blob. After that, slowing down displacement of the
positron, $R_{sd}(W_{bl},{\rm Ry}) - a_{bl}$ should be equal to the "radius" of the blob, $a_{bl}$, i.e. the
blob in average embraces exactly the end part of the e$^+$ ionization slowing down trajectory.

Solution of these equations is unique and shown in Fig.\ref{a_bl_W_bl} for the case of liquid water. It is
seen that numbers $W_{bl}\approx 500$ eV and $a_{bl}=40$ \AA\ fulfill Eqs.(A4). Actually, we must proceed
with all such calculations for each particular liquid investigated by the positron spectroscopy. However, one
may assume that the values of $a_{bl}$, $W_{bl}$ and other parameters related to slowing by ionization do not
differ significantly from one liquid to another, because variations in the ionization potential and average
electron density are small (Table I).

At the end of slowing down by ionization and electronic excitation, the spatial distribution of
the subionizing positron coincides with the distribution of the blob species (i.e. $\sim \exp
(-r^2/a_{bl}^2)$). Further, during thermalization, e$^+$ distribution becomes broader (about 30
\AA\ in water and 100-200 \AA\ in hydrocarbons), and its total dispersion is expressed as
follows:
$$
a_p^2 \approx a_{bl}^2 + 2 \left\langle \int_{T}^{W_0} l_{tr}(W) {dW \over |dW/dx|_{vib}} \right\rangle_{W_0}. \eqno(A5)
$$
Estimation of $a_p$ requires knowledge of the stopping power, $|dW/dx|_{vib}$, of a given liquid towards excitation of
vibrations, scattering and energy loss properties of subionizing e$^+$, and the spectrum of its initial energies $W_0$ after
the last ionization event. $\langle \dots \rangle_{W_0}$ denotes the average over $W_0$. \cite{Bya96,Ste95} Thus, contrary to
the parameters related to ionization slowing down, $a_p$ strongly depends on the properties of each particular liquid.

\section{Microscopic quasi-equilibrium condition for ions and electrons in the blob}

Let us estimate how much the distributions of electrons and ions differ at quasi-neutrality
(quasi-equilibrium) condition which is achieved immediately after the electrons adjust themselves
to the current distribution of ions.  As discussed before Eq.(3), it is reasonable to use
Gaussian distribution functions to describe the spatial distribution of ions and electrons by the
end of the thermalization stage:

$$
n_0 G_i (r) = n_0 {\exp(-r^2/a_i^2) \over \pi^{3/2} a_i^3}
    \qquad
n_0 G_e (r)= n_0 {\exp(-r^2/a_e^2) \over \pi^{3/2} a_e^3}.
\eqno(B1)
$$
$n_0$ is the number of ion-electron pairs in the blob.  We expect that, due to strong
electrostatic attraction, the difference $\Delta a = a_e - a_i$ is very small in comparison with
the blob radius $a_{bl} \approx a_i \approx a_e$. The distribution of electrons is slightly
broader than that of ions because electrons are much more mobile.

Integrating the equation on electric field ${\bf E}(r)$

$$
\hbox{div}{\bf E}(r)  = {4\pi e n_0 \over \varepsilon} (G_i-G_e),
\eqno(B2)
$$
we find the electric field, produced by electrons and ions in a liquid:

$$
{\bf E}(r)  = {2\over \sqrt{\pi}} {e n_0 \over \varepsilon r^2}
              {{\bf r} \over r}
              \int_{(r/a_e)^2}^{(r/a_i)^2} \sqrt{x} \exp(-x)dx
    \stackrel{\Delta a \ll a_{bl}}{\longrightarrow}
    \frac{4}{\sqrt{\pi}}    \frac{e n_0}{\varepsilon a^2_{bl}}
    \frac{\Delta a}{a_{bl}} \frac{{\bf r}}{a_{bl}}
    \exp\left(-\frac{r^2}{a_{bl}^2} \right).
\eqno(B3)
$$
At the quasi-equilibrium, the electron flux should equal to zero everywhere. However, here we use
the approximate (gaussian) functions instead of true distribution functions, so we can fulfill
this condition only for a given value of $r$, for example, at $r=a_{bl}$:

$$
c_e b_e {\bf E}(a_{bl}) + D_e \nabla c_e(a_{bl}) = 0,
    \qquad
c_e(a_{bl}) = n_0 G_e(a_{bl}),
    \qquad
b_e = {e D_e \over T}.
\eqno(B4)
$$
From this equation in the limit $a_e - a_i \ll a_{bl}$, we find difference $a_e - a_i$. Using
Eq.(B1) and Eq.(B3), we obtain
$$
\nabla c_e (a_{bl}) = - {2 \over a_{bl} } c_e(a_{bl}) {{\bf r}\over r},
    \qquad
{\bf E}(a_{bl}) = \frac{4}{\sqrt{\pi}{\rm e}} \cdot {e n_0 \Delta a \over \varepsilon a_{bl}^3}
{{\bf r}\over r},
    \qquad
{\rm e} = 2.718 \ldots ~.
\eqno(B5)
$$
Substituting these equations into Eq.(B4), we arrive at
$$
\Delta a \simeq {2.4 a_{bl}^2 \over n_0 r_c }, \qquad r_c = {e^2 \over \varepsilon T}. \eqno(B6)
$$
Numerically, it yields $\Delta a \simeq 1$ \AA\ (an estimation of $n_0$ and $a_{bl}$ is given in Appendix A).  The potential,
which corresponds to Eq.(B3), is
$$
\varphi (r) =
    \frac{2}{\sqrt{\pi}} \frac{e n_0}{\varepsilon a_{bl}}
     \frac{\Delta a}{a_{bl}} \exp (-
             r^2/a_{bl}^2) =
    \frac{e}{\varepsilon r_c} \exp (1-
r^2/a_{bl}^2), \qquad
-\nabla \varphi (r) = {\bf E}(r).
\eqno(B7)
$$

Note that the application of an external field of about 30 kV/cm to a liquid does not significantly perturb the ion-electron
distribution in the blob. It leads to less than 1 \AA\ shift of the electron distribution with respect to the ions.

\section{Ambipolar diffusion expansion of the blob}

Even neglecting ion-electron recombination, the out-diffusion of the blob species (electrons and
ions) can not be considered independently because of the strong electrostatic interaction between
charged particles.  The intrablob electric field efficiently suppresses out-diffusion of the
electrons but increases (duplicates) the ion diffusion coefficient. \cite{Lif81} Coupled diffusion
equations are written as

$$
{\partial c_i({\bf r},t) \over \partial t} =
   D_i \left( \Delta c_i + {c_i e \over T} \Delta \varphi
\right), \eqno(C1a)
$$
$$
{\partial c_e({\bf r},t) \over \partial t} =
   D_e \left( \Delta c_e - {c_e e \over T} \Delta \varphi
\right). \eqno(C1b)
$$
$D_i$ and $D_e$ are the local diffusion coefficients of ions and electrons and $\Delta$ stands
for the Laplace operator.  Our aim now is to relate them to the kinetics of blob expansion.  As
before, Eqs.(3) are used as initial conditions.  To Eqs.(C1), we add the Poisson equation for an
electrostatic potential $\varphi(r)$:

$$
\Delta \varphi = -{4\pi e \over \varepsilon} (c_i - c_e).
\eqno(C2)
$$
In Appendix A, we demonstrate that the spatial distribution of electrons is very close to that
of ions, so  $\delta c = c_i - c_e \ll c_i$ or $c_e$. Using Eq.(C2), we evaluate the second
terms in the right hand side of Eq.(C1) as follows:

$$
{c e \over T} \Delta \varphi \sim 4\pi c r_c \delta c,
\qquad
c(r) \equiv c_e \approx c_i,
\qquad
r_c = {e^2 \over \varepsilon T}.
$$

$\Delta c_e$ and $\Delta c_i$ are on the order of $c/a_{bl}^2$.  Their difference, $\Delta c_e
-\Delta c_i \sim \delta c /a_{bl}^2$, is negligible compared to the second term, $4\pi c r_c
\delta c$, on the right hand side of Eqs.($C$1), because their ratio is small:

$$
{\delta c /a_{bl}^2 \over 4 \pi c r_c \delta c } =
    { a_{bl} \over 4 \pi n_0 r_c} \ll 1.
$$
We assumed $c\sim n_0/a_{bl}^3$.

Moving fast ($D_e \gg D_i$), electrons rapidly adjust themselves to the current distribution of
ions. So $\partial c_e/ \partial t =0$, which gives

$$
{c_e e \over T} \Delta \varphi = \Delta c_e.
\eqno(C3)
$$
Together with $\Delta c_e \approx \Delta c_i$ and ${c_i e \over T} \Delta \varphi \approx{c_e e
\over T} \Delta \varphi $, we arrive at

$$
{c_i e \over T} \Delta \varphi = \Delta c_i.
$$
This means that the term $\Delta c_i$ is duplicated in Eq.(C1$a$). The motion of ions in the
blob thus obeys the simple diffusion equation, but with a twice larger diffusion coefficient:

$$
{\partial c_i({\bf r},t) \over \partial t} = 2D_i \Delta c_i.
\eqno(C4)
$$
This process is called ambipolar diffusion and $D_{amb}=2D_i$ is the ambipolar diffusion
coefficient. Half of it is related to the diffusion of ions and the rest is due to the electric
field of blob electrons.

\clearpage
\centerline{Figure Captions}

\bigskip
\bigskip
\noindent
Figure 1
\medskip

\noindent Dependence of o-Ps intensity $I_3 = 3P_{\rm Ps}/4\cdot 100\%$ on external electric field D in
different liquids: $\square$   -- cyclohexane (C$_6$H$_{12}$), $\oplus$  -- n-hexane (C$_6$H$_{14}$),
$\triangle$ -- isooctane (C$_8$H$_{18}$), {\Large $\diamond$} -- hexafluorobenzene (C$_6$F$_6$) and {\Large
$\circ$} -- benzene (C$_6$H$_6$). The size of symbols reflects a statistical uncertainty only.  Broken lines
represent the fit according to Eq.(22), which does not account for the free positron annihilation. They
solely show the limiting case of the theory. Solid lines show the fit when positron annihilation with the
lifetime $\tau_2$ and IER ($W_{ie} / W_{ep} = 0.003$) are taken into account in Eq.(32). The lines which
account only effect of the positron annihilation lie in between respective dashed and solid lines and not
shown here. Experimental data suggests the presence of IER on the level $W_{ie} / W_{ep} < 0.005$ or positron
localization with a rate lower than $1/\tau_2$.  Corresponding parameter values are shown in Table II.

\bigskip
\bigskip
\noindent Figure 2
\medskip

\noindent Scheme of the end part of the e$^{+*}$ track and Ps formation.

\bigskip
\bigskip
\noindent Figure 3
\medskip

\noindent Dependence of the mean distance between adjacent ionizations $l_i(W)$ and the
transport mean free path $l_{tr}(W)$ of the positron on its kinetic energy. \cite{Bya96} Spurs
start to overlap when $l_i(W_{cyl})\lesssim 2a_{sp}$, forming cylindrical-like column (the arrow
at right). Formation of the blob begins when the positron energy fulfills equation (A4):
$l_{tr}(W_{bl})=R_{sl}(W_{bl},Ry)/2$.

\bigskip
\bigskip
\noindent Figure 4
\medskip

\noindent Schematic view of the terminal positron blob. Positron motion is simulated as random
walks with the energy dependent step, $l_{tr}(W)$.

\clearpage
\noindent
\vspace{2cm}

\clearpage
\noindent

\clearpage
\noindent
\vspace{2cm}

\clearpage \noindent \vspace{2cm}

\clearpage
\begin{table}
\caption{Parameters of the liquids}
\begin{center}
\begin{tabular}{|c|c|c|c|c|c|c|c|c|c|c|c|}
\hline
                 &$R_{\rm WS}$
                       &$\varepsilon$
                             &$Z$
                                &$n_e$
                                      &$I$
                                            &$\tau_3$
                                                    &$\tau_2$
                                                          &$b_e$
                                                                   &$b_p$
                                                                         &$r_0$
                                                                         &$V_0^-$
                                                                             \\
                  &\AA &     &  &\AA$^{-3}$
                                      & eV
                                                 & ns
                                                         & ns
                                      &${\hbox{{\footnotesize cm$^2$}} \over
                                      \hbox{{\footnotesize V$\cdot$s}}}$
                                      &${\hbox{{\footnotesize cm$^2$}} \over
                                      \hbox{{\footnotesize V$\cdot$s}}}$
                                                                         &\AA
                                                                         &eV
                                                                             \\
\hline
hexafluoro-      &3.57 &2.03 &90&0.471&9.89$^a$&3.64(1)&0.61(3)&0.011&0(7)$^c$   & 52 &         \\
benzene          &     &     &  &     &         &       &       &     &          &    &        \\
C$_6$F$_6$       &     &     &  &     &         &       &       &     &          &    &        \\
\hline
isooctane        &4.04 &1.96 &66&0.239&9.86$^a$ &4.16(2)&0.53(3)&4.5  &134(6)$^c$&110 &$-0.35$ \\
i-C$_8$H$_{18}$  &     &     &  &     &8.4$^b$  &       &       &     &          &    &        \\
\hline
n-hexane         &3.74 &1.89 &50&0.229&10.13$^a$&4.01(2)&0.55(3)&0.07-&100(6)$^c$&60- &$-0.07$-\\
n-C$_6$H$_{14}$  &     &     &  &     &8.7$^b$  &       &       &0.09 &100$^d$   &67  &0.1     \\
\hline
cyclohexane      &3.50 &2.02 &48&0.267&9.86$^a$ &3.28(2)&0.50(4)&0.24-&          &59- &0.01    \\
c-C$_6$H$_{12}$  &     &     &  &     &8.4$^b$  &       &       &0.45 &          &67  &        \\
\hline
benzene          &3.28 &2.27 &42&0.283&9.24$^a$ &3.20(2)&0.55(4)&0.11-&8.4$^d$   &42  &$-0.5$- \\
C$_6$H$_6$       &     &     &  &     &7.1$^b$  &       &       &0.14 &          &    &$-0.14$ \\
\hline
\end{tabular}
\end{center}
$R_{\rm WS}$ is the radius of the Wigner-Seitz sphere at room temperature, $4\pi R_{\rm WS}^3
/3 = 1/n$, where $n$ is the molecular concentration.\\
$\varepsilon=\varepsilon_\infty$ is the dielectric permittivity of the liquid. \cite{HCP98}\\
$Z$       is the total number of electrons in molecule \\
$n_e=Zn$: Average electron density of liquid \\
$\tau_3$: Ortho-Ps lifetime. \cite{Ste00}\\
$\tau_2$: Free-e$^+$ lifetime. \cite{Ste00}\\
$b_e$:    Mobility of an excess electron. \cite{HRC91,Gerin} \\
$b_p$:    Positron mobility. In \cite{Wan00} e$^+$ mobility was defined as a coefficient
          of proportionality between e$^+$ drift velocity and external electric field D. Thus, in our notation
          respective numbers should be multiplied by a factor of $\varepsilon$.\\
$r_0$:    Ion-electron initial separation in a spur. \cite{HRC91,Gerin}\\
$V_0^-$:  Electron work function. \cite{HRC91} \\
$^{a)}$   Gas-phase ionization potential ($I_G$). \cite{HCP98} \\
$^{b)}$   Liquid phase ionization potential ($I_L$). \cite{Cas80} \\
$^{c)}$   Reference \cite{Wan00}\\
$^{d)}$   Reference \cite{Ito88}\\

\end{table}

\clearpage
\begin{table}
\caption{Parameters obtained from the fit of the data}
\begin{center}
\begin{tabular}{|c|c|c|c|c|c|}
\hline
                 &$(a_{bl}^2 +a^2_p)^{1/2}$
                         &$W_{ep}$&$W_p$ &$k_{ep}$           &$k_{ie}$           \\
                 & \AA   &        &      &M$^{-1}$s$^{-1}$   &M$^{-1}$s$^{-1}$   \\
\hline
hexafluoro-      &43(3)  &0.642(3)& -    &$3.1\cdot 10^{13}$ &-                  \\
          benzene&59(3)  &0.670(3)&0.0006&$4.5\cdot 10^{13}$ &-                  \\
C$_6$F$_6$       &68(4)  &0.687(3)&0.0015&$5.3\cdot 10^{13}$ &-                  \\
                 &79(4)  &0.709(3)&0.0010&$6.3\cdot 10^{13}$ &$7.0\cdot 10^{10}$ \\
\hline
isooctane        &153(3) &0.411(3)& -    &$9.4\cdot 10^{14}$ &-                  \\
i-C$_{8}$H$_{18}$&160(3) &0.419(3)&0.0004&$1.0\cdot 10^{15}$ &-                  \\
                 &164(4) &0.422(3)&0.0008&$1.0\cdot 10^{15}$ &-                  \\
                 &173(4) &0.431(3)&0.0004&$1.1\cdot 10^{15}$ &$1.2\cdot 10^{11}$ \\
\hline
n-hexane         &161(3) &0.379(3)& -    &$6.8\cdot 10^{14}$ &-                  \\
n-C$_6$H$_{14}$  &171(3) &0.388(3)&0.0005&$7.4\cdot 10^{14}$ &-                  \\
                 &177(4) &0.394(3)&0.0011&$7.8\cdot 10^{14}$ &-                  \\
                 &185(4) &0.400(3)&0.0006&$8.3\cdot 10^{14}$ &$7.1\cdot 10^{10}$ \\
\hline
cyclohexane      &196(4) &0.326(3)& -    &$7.7\cdot 10^{14}$ &-                  \\
c-C$_6$H$_{12}$  &212(4) &0.336(3)&0.0008&$8.5\cdot 10^{14}$ &-                  \\
                 &222(5) &0.343(3)&0.0018&$9.1\cdot 10^{14}$ &-                  \\
                 &226(5) &0.344(3)&0.0010&$9.3\cdot 10^{14}$ &$4.4\cdot 10^{10}$ \\
\hline
benzene          &166(3) &0.410(3)& -    &$1.5\cdot 10^{14}$ &-                  \\
C$_6$H$_6$       &214(5) &0.451(3)&0.0043&$2.1\cdot 10^{14}$ &-                  \\
                 &242(5) &0.484(3)&0.0112&$2.5\cdot 10^{14}$ &-                  \\
                 &224(5) &0.461(3)&0.0048&$2.2\cdot 10^{14}$ &$1.1\cdot 10^{10}$ \\
\hline
\end{tabular}
\end{center}

\medskip
The first line for each liquid corresponds to the fit based on Eq.(22) (dashed lines in Fig.\ref{I3_D}),
which does not account for free positron annihilation. We list these numbers to show the simplest limiting
case of the theory. These numbers differ from corresponding data in \cite{Ste00} because the field-dependent
bias of subionizing positron during its thermalization is taken into account in the present work.

Numbers on the second line show the fit including positron annihilation with lifetime $\tau_2$
based on Eqs.(32-33), but IER was not taken into account.

Data on the third line corresponds to the same case as above, but positron localization is assumed with the
rate $1/\tau_p^{loc} = 1/\tau_2$. Respective curves for $P_{\rm Ps}(D)$ in the last two cases are not shown
in Fig.\ref{I3_D}. They are in between the dashed and the solid lines.

Numbers on the last line of each group are obtained based on Eq.(32), where, together with the positron
annihilation with the rate $\tau_2^{-1}$, a fixed fraction of IER is introduced:  $W_{ie} / W_{ep}= 0.003$
(maximum possible level still acceptable by experimental data).

$k_{ep}$ and $k_{ie}$ are recalculated from $\sqrt{a_{bl}^2+a_p^2}$, $W_{ep}$  and $W_{ie}$ via Eq.(21) and
Eq.(31). It is assumed that $n_0 = 30$, $a_{bl}= 40$ \AA\ and $D_p = T b_p/e$, where $b_p$ is taken from
Table I (e$^+$ mobility for cyclohexane is assumed to be the same as for hexane).

\end{table}

\end{document}